# Contrastive Learning of Subject-Invariant EEG Representations for Cross-Subject Emotion Recognition

Xinke Shen†, Xianggen Liu†, Xin Hu, Dan Zhang, *Member, IEEE* and Sen Song

**Abstract**—EEG signals have been reported to be informative and reliable for emotion recognition in recent years. However, the inter-subject variability of emotion-related EEG signals still poses a great challenge for the practical applications of EEG-based emotion recognition. Inspired by recent neuroscience studies on inter-subject correlation, we proposed a Contrastive Learning method for Inter-Subject Alignment (CLISA) to tackle the cross-subject emotion recognition problem. Contrastive learning was employed to minimize the inter-subject differences by maximizing the similarity in EEG signal representations across subjects when they received the same emotional stimuli in contrast to different ones. Specifically, a convolutional neural network was applied to learn inter-subject aligned spatiotemporal representations from EEG time series in contrastive learning. The aligned representations were subsequently used to extract differential entropy features for emotion classification. CLISA achieved state-of-the-art cross-subject emotion recognition performance on our THU-EP dataset with 80 subjects and the publicly available SEED dataset with 15 subjects. It could generalize to unseen subjects or unseen emotional stimuli in testing. Furthermore, the spatiotemporal representations learned by CLISA could provide insights into the neural mechanisms of human emotion processing.

---◆---

## 1 INTRODUCTION

Electroencephalogram (EEG) based emotion recognition has gained increasing interest in recent years [1], [2]. Unlike behavioral techniques that record facial expressions, body gestures, voice, etc. [3], EEG provides a more direct and objective measurement of human emotional responses that cannot be easily disguised or consciously restrained [1], [4], [5]. Compared to other neuroimaging techniques such as functional magnetic imaging (fMRI) and magnetoencephalography (MEG), EEG is advantageous for its portability and cost-effectiveness in real-world applications [4], [5].

Extensive studies have investigated emotion-related EEG representations [6], [7], [8]. Differential entropy (DE) features, which are equivalent to the logarithm energy spectrum in a specific frequency band [9], [10], have been widely used in the state-of-the-art emotion recognition methods [11], [12], [13]. Beta- and Gamma-band DE features in temporal regions were found to be closely related to emotion [14], [15]. Beyond localized DE features, researchers found that modeling the relationships among EEG electrodes was critical for emotion recognition. In this direction, graph neural networks (GNNs) [13], [16], [17] and long short-term memory (LSTM) [11], [12] were proposed to extract spatial relationships of DE features among different EEG channels. Network-based features like phase locking values [18] and microstate parameters [19] were also proposed to extract the coactivation patterns of different brain regions directly. Furthermore, convolutional neural networks (CNNs) and attention mechanisms were utilized to learn the emotion-related EEG representations in an end-to-end manner [20], [21], [22]. These methods employed the prominent representational power of deep neural networks to avoid human-crafted feature extraction.

However, most studies have mainly focused on intra-subject emotion recognition [5]. To reach satisfactory recognition performance, researchers needed to collect sufficient data (generally half an hour to more than one hour) within one subject to learn subject-dependent EEG representations for emotion recognition [14], [23], [24]. This laborious and time-consuming training procedure has become a major bottleneck for the practical use of EEG-based emotion recognition. Therefore, developing emotion recognition methods with good cross-subject generalizability is desirable for realistic applications, especially in the cases of new users [25], [26].

The substantial inter-subject variabilities of emotion-related EEG activities posed great challenge for cross-subject emotion recognition [27], [28]. For example, the cross-subject emotion recognition accuracy on the widely used SEED dataset [9], [14] could be as low as 58% with a generic multiple layer perceptron classifier. In comparison, the

• †These authors contributed equally to this work.
• X. Shen and S. Song are with the Department of Biomedical Engineering and with the Tsinghua Laboratory of Brain and Intelligence, Tsinghua University, Beijing, China, 100084. Email: sxk17@mails.tsinghua.edu.cn; songsen@tsinghua.edu.cn.
• X. Liu is with the College of Computer Science, Sichuan University, Chengdu, China, 610065. E-mail: liuxianggen@scu.edu.cn.
• X. Hu and D. Zhang are with the Department of Psychology and with the Tsinghua Laboratory of Brain and Intelligence, Tsinghua University, Beijing, China, 100084. E-mail: huxin530@gmail.com; dzhang@tsinghua.edu.cn.
• D. Zhang and S. Song are the corresponding authors.

intra-subject emotion recognition could achieve high accuracy of 96% with the same classifier [29]. The substantial drop in the emotion recognition performance from the intra-subject to the cross-subject scenario could be explained by the well-acknowledged individual difference in EEG-based emotion representations due to factors such as individualized experience and dispositional characteristics [30], [31], [32], [33]. Nonetheless, subject-invariance of emotion representation has also been well documented in the field of psychology and neuroscience [15], [34], [35], [36], [37]. Previous fMRI and EEG studies have identified distinct and stable neural representations for different emotions across subjects [15], [34], [35], [36], [37], suggesting the possibility of developing cross-subject emotion recognition algorithms.

To address the issue of inter-subject variability, researchers have applied domain adaptation (DA) and domain generalization (DG) methods to cross-subject emotion recognition [26], [38]. The DA methods aim to minimize the discrepancy between data distributions of the source domain (i.e., the training subject) and the target domain (i.e., the testing subject). These methods have to access data from the target domain during the training process to measure the data discrepancy. As an example of the DA methods, domain-adversarial neural networks (DANNs) leverage adversarial training to align the EEG representation of the source domain and the target domain [39]. It was adopted by multiple state-of-the-art cross-subject emotion recognition models [11], [12], [13] and could improve the accuracy from approximately 60% to more than 80% on the SEED dataset. The DG methods, on the other hand, find the domain-invariant representations from the source domains. In contrast to DA, it does not need to access the testing subjects' data, so it is preferred in real-world applications. An adversarial domain generalization method recently achieved comparable results with the DA methods on the SEED dataset [26]. The success of these methods indicated the possibility of finding subject-invariant EEG representations for emotion recognition. However, the cross-subject emotion recognition methods to date have been developed mainly from the machine learning perspective, with very limited consideration of the neuroscientific basis of human emotion processing.

The emerging neuroscience studies on inter-subject correlation (ISC) could offer a new perspective for exploring subject-invariant emotion representations and developing cross-subject emotion recognition methods [40], [41], [42], [43], [44], [45], [46]. The ISC approach, originally proposed to investigate the perception of naturalistic visual scenes, focuses on the synchronization of neural activities (e.g., EEG) between subjects when perceiving the same stimuli. Taking this inter-subject perspective, the temporal, spatial, and spectral patterns of ISC could reveal the neural mechanisms of information processing for naturalistic stimuli such as movies and narrative speech [40], [41], [43]. In the field of emotion, several pioneering studies have shown that the ISC of EEG signals among a group of subjects watching the same emotional videos could reflect their group-level preference, arousal, valence, etc. [4], [41], [42]. These findings suggest that the stimulus-specific EEG responses shared across individuals could carry valuable information for discriminating different emotional states. More importantly, the effectiveness of the shared EEG responses provides critical neuroscience evidence in favor of constructing subject-invariant emotion representations. Nonetheless, it remains elusive how to extract the subject-invariant emotion representations effectively and make them generalizable to new subjects and new stimuli.

In this work, we propose a data-driven approach that performs Contrastive Learning for Inter-Subject Alignment (CLISA). Inspired from the neuroscientific observations of ISC, CLISA is grounded on the assumption that the neural activities of the subjects are in a similar state when they receive the same segment of emotional stimuli (i.e., the emotional videos in our study). Based on this fundamental idea, we propose to learn a subject-invariant space for EEG signals by aligning the representations underlying similar mental activities. Specifically, our CLISA framework contains two phases, i.e., the contrastive learning procedure and the prediction procedure. In the contrastive learning procedure, a convolutional neural network (CNN)-based encoder learns invariant and predictive spatiotemporal representations of EEG signals. It maximizes the similarities of the representations in response to identical emotional stimuli (positive pairs) while minimizing the similarities between signals corresponding to different stimuli (negative pairs). In the prediction procedure, a classifier together with the trained encoder takes EEG signals as inputs to identify human emotions. Considering that the contrastive pairs can be easily constructed from EEG datasets [14], [23], [24] and they are of great number in the contrastive learning procedure, the learned representations are expected to be informative and generalizable in the prediction procedure.

The advantages of CLISA are three-fold: 1) CLISA virtually and vastly increases the training data for the learning of EEG representations by contrastive learning. 2) CLISA can generalize to new subjects without requiring extensive data from them, thus enhancing the practicality of emotion recognition systems. 3) Benefiting from the contrastive learning strategy and the considerably larger amount of training samples, the learned representations are not only invariant to subjects but also generalizable to different stimuli. That is to say, the representation is not specific to the particular stimuli used in training. Instead, it is a general representation for emotion processing.

## 2 RELATED WORK

### 2.1 EEG-based Cross-subject Emotion Recognition

Domain adaptation (DA) has been demonstrated as an effective technology in cross-subject emotion recognition. It aims to deal with the domain shift problem during training and testing. For the classical DA methods, Zheng & Lu [47] compared the performance of transfer component analysis (TCA) [48], kernel principal component analysis (KPCA) [49], transductive parameter transfer (TPT), etc. on the SEED dataset [50]. TPT achieved the best accuracy of 76.3%, with a considerable improvement of 19.6% over the generic classifier with no DA. Further, Chai et al. [51]

proposed an adaptive subspace feature matching (ASFM) strategy. Its essential component is a subspace alignment (SA) algorithm [52], which linearly transforms the PCA subspace of the training subject's data to be aligned with that of the testing subject's data.

In addition to these classical DA methods, DA with deep learning has also been utilized in cross-subject emotion recognition. In particular, Chai et al. [53] proposed an auto-encoder architecture to reduce the discrepancy of training and testing subjects in a learned latent space. Domain-adversarial neural networks (DANNs) employed a domain classifier and a gradient reversal layer to enforce the network to learn domain-indiscriminate representations across the source domain and the target domain [39]. Based on this strategy, researchers further proposed the bi-hemispheres domain-adversarial neural network (Bi-DANN) [11] and regularized graph neural network (RGNN) [13] to facilitate cross-subject emotion recognition. BiDANN contained two local domain classifiers for each hemisphere and a global domain classifier to learn subject-invariant emotion representations. RGNN introduced a node-wise domain-adversarial training in graph neural networks, which was better than graph-level domain-adversarial training. In addition to DANN, Li et al. [29] proposed a joint domain adaptation model to align the marginal and conditional distribution of the data simultaneously. These methods have improved the performance of cross-subject emotion recognition to around or above 85% on the SEED dataset, demonstrating their superiority to most of the classical DA methods. However, all the methods above had to access extensive data from the testing subjects for domain-adversarial training. Similar to intra-subject emotion recognition, half-an-hour to one-hour data from a new subject (although no requirements for emotional labels) were needed in general, which still hindered the application of emotion recognition methods in the real world.

In recent years, domain generalization (DG) methods have been applied to cross-subject emotion recognition to alleviate the reliance on data from testing subjects. DG models are trained to extract the domain-invariant representations across multiple training subjects, and they are ready to be applied to new subjects without access to their data. For example, Ma et al. [26] built a domain residual network to learn domain-shared weights and domain-specific weights with domain-adversarial training. Then the domain-shared weights were used to classify emotions for unseen subjects. Besides, Zhao et al. [54] proposed an autoencoder architecture to learn domain-shared encoders and classifiers, which showed generalizability to a new subject using his/her data within only one minute for calibration. These methods have achieved similar performance with DA methods on the SEED dataset.

## 2.2 Contrastive Learning

Contrastive learning is a kind of self-supervised learning algorithm that learns to discriminate whether pairs of data are similar or not. It has achieved state-of-the-art performance in various fields such as computer vision [55], natural language processing (NLP) [56], and bioinformatics [57], [58]. Contrastive learning can be generally divided into two types: 1) context-instance contrast, or global-local contrast, such as assigning a sentence to its paragraph and associating strides to a zebra, and 2) instance-instance contrast, like identifying a transformed image with the original one [59]. After the pretext contrastive learning, the model can generate better data representations for downstream tasks.

For EEG, a few studies have used contrastive learning methods to learn data representations from relatively large datasets and then applied the pretrained model to downstream tasks. Mohsenvand et al. [60] enforced the model to learn similarities between different views of augmented samples from the same original data, similar to the popular SimCLR framework (a simple framework for contrastive learning of visual representations) [55]. The samples were augmented by temporal masking, linear scaling, Gaussian noise adding, etc. The model pretrained on the combination of three datasets performed well on multiple downstream tasks: sleep stage classification, clinical abnormal detection, and emotion recognition. Banville et al. [61] used temporal context prediction and contrastive predictive coding [62] to learn data representations on two clinical datasets. They demonstrated that the pretrained model could extract latent structures of age effects, gender effects, and pathological effects. These methods generally adopted the contrastive learning framework from computer vision or natural language processing directly. Benefitted from the contrastive learning strategy and large datasets in pretraining, they could denoise the data or learn the temporal dependencies of the data. In this work, we proposed a contrastive learning strategy specifically designed for cross-subject generalization. It learns the similarity between samples from different subjects when they were presented with the same stimuli. Our method does not need extensive external data but creates large self-supervised labels based on the inter-subject aligned experimental design.

As our contrastive learning strategy was inspired by inter-subject correlation (ISC) studies, we will introduce the recent neuroscience findings of ISC here. The ISC studies have focused on the inter-subject consistent neural activities in response to the same naturalistic stimuli [40], [41], [42], [43] or in the same social interaction scenarios [44], [45]. They provided insights into how human perceptions and emotions can be shared and why human communications can be effective from the neuroscience perspective. In a keynote fMRI study, Hasson et al. [43] found voxel-by-voxel synchronization across subjects when they were watching the same movie, revealing that different brains tend to respond similarly to the same naturalistic stimuli. Later, Dmochowski et al. discovered that the level of ISC in EEG signals could reflect people's emotionally laden attention to the movie [42] and predict the preference for the video clips in a large population [41]. Ding et al. [4] further reported that EEG-based ISC could effectively predict real-time emotional experiences of arousal and valence. The prediction performance increased as a function of the number of subjects (i.e., brains) included in the analysis. These findings supported that the invariant response across individuals could be informative about various mental states, serving as solid theoretical supports for our contrastive learning strategy. However, due to the complexity of neural signals like EEGs, current ISC analysis has emphasized that the subjects should be positioned in the same sensory environment (and preferably engaged in similar tasks) to facilitate the computation of ISCs [46], [63]. How to extract inter-subject invariant representations that can generalize to new subjects or new stimuli has rarely been explored.

Fortunately, the paradigms adopted by the mainstream emotion EEG datasets, such as SEED [14], DEAP [23], and DREAMER [24], have made them well suited for our contrastive learning method. Specifically, the data were generally collected from a group of subjects undergoing the same task with the same emotional stimuli. Therefore, contrasting the EEG signals in response to the same or different emotional stimuli from two subjects would be helpful to learn the EEG representations for emotion processing that are shared across subjects. If the learned representation could capture the subject-invariant emotional EEG responses, it would be expected to generalize well to unseen subjects and unseen emotional stimuli to predict the emotional states.

Last but not least, the contrastive learning framework would benefit from increasing the number of subjects. The number of contrasts could increase quadratically with the increasing number of subjects. However, to our knowledge, the number of subjects usually ranged from 10 to 40 in the publicly available emotional EEG datasets [14], [23], [24]. In the present study, in addition to the popular SEED dataset with 15 subjects, a new dataset THU-EP (the Tsing-Hua University Emotional Profile dataset [64]) with 80 subjects was included (see Section 4.1 for more details), which was expected better to evaluate the effectiveness of the proposed CLISA method.

## 3 THE CONTRASTIVE LEARNING METHOD FOR INTER-SUBJECT ALIGNMENT (CLISA)

This section presents our *Contrastive Learning method for Inter-Subject Alignment* (CLISA) (Fig. 1). We first introduce the contrastive learning procedure and then explain the prediction procedure of CLISA. In the contrastive learning procedure, we utilize a base encoder and a projector with convolutional layers to align the data representations across individuals. In the prediction procedure, we use the learned representations to identify the emotion labels of the EEG signals.

### 3.1 The Contrastive Learning Procedure

In the contrastive learning procedure, CLISA contains four components: a data sampler, a base encoder, a projector, and a contrastive loss function. First, the data sampler generates a minibatch containing several pairs of EEG segments for training. Next, the base encoder processes these segments using a spatial convolution operator and a temporal convolution operator, aiming to transform the data from individual subjects to inter-subject aligned representations. Then the projector maps the representations to another latent space for computing their similarities (Fig. 1). Together, the parameters of the base encoder and projector are optimized to minimize contrastive loss.

#### A. The Data Sampler

In our contrastive learning strategy, the model learns to distinguish whether two series of EEG signals correspond to the same stimuli (i.e., the same segment of the video). To achieve this goal, we design a data sampler to prepare the inputs in a minibatch manner. In the EEG datasets, the data from one subject consist of $N$ trials. In each trial, the subject watched one emotional video. To obtain a minibatch, we firstly randomly extract one EEG sample from each trial of subject $A$ (the time length of one sample is smaller than that of one trial), resulting in $N$ samples. We denote them as $X_1^A, X_2^A, \dots, X_N^A$ ($X_i^A \in \mathbb{R}^{M \times T}$, $M$ is the number of EEG channels and $T$ is the number of time points in one EEG sample). Then we extract $N$ samples of EEG signals from another subject $B$ similarly. We denote them as $X_1^B, X_2^B, \dots, X_N^B$. The samples $X_i^A$ and $X_i^B$ correspond to the same time segment of trial $i$ ($i = 1, 2, \dots, N$). The set $\mathcal{D} = \{X_i^s | i = 1, 2, \dots, N; s \in$ $\{A, B\}\}$ constitutes one minibatch (Fig. 2). In this minibatch, given a sample $X_i^A$, the sample $X_i^B$ forms a positive pair with it, and the other samples $\{X_j^s | j = 1, 2, \ldots, N, j \neq i; s \in \{A, B\}\}$ form $2(N-1)$ negative pairs with $X_i^A$. We enumerate all possible subject pairs $\{A, B\}$ in the training set for one contrastive learning epoch. For example, if there are 15 subjects in the dataset, the number of subject pairs is $15 \times 14 / 2 = 105$. If there are 80 subjects, the number of subject pairs is $80 \times 79 / 2 = 3160$.

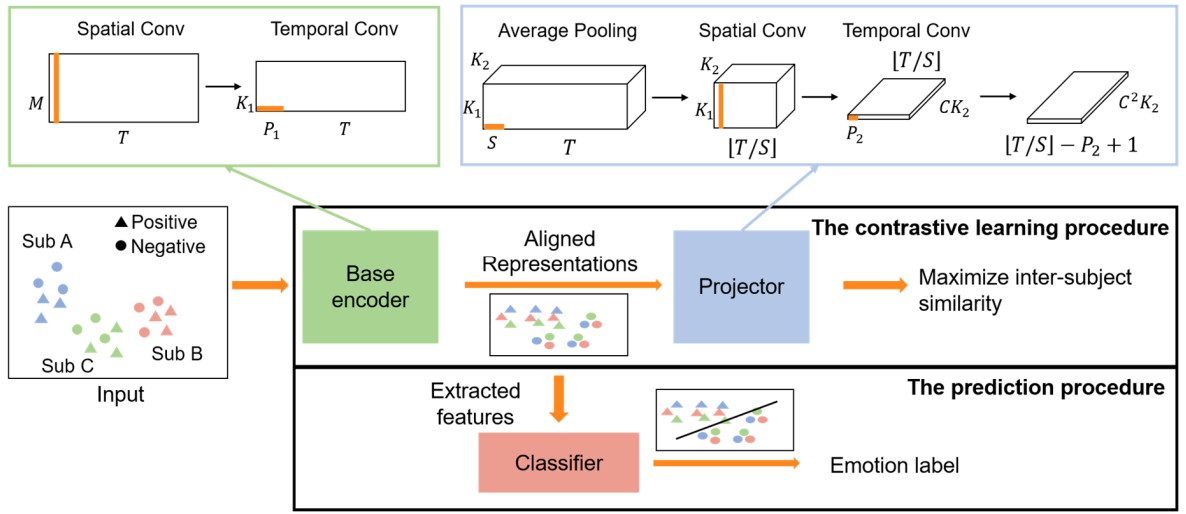

Fig. 1. The illustration of the Contrastive Learning method for Inter-subject Alignment (CLISA). In the figure, "Sub" stands for "subject" and "Conv" stands for "Convolution." Upper-left: The architecture of the base encoder. Upper-right: The architecture of the projector. Lower: The procedures of contrastive learning and prediction.

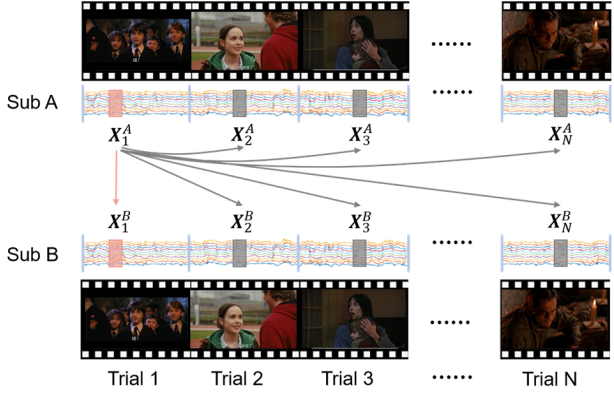

Fig. 2. The illustration of the data sampler. In a minibatch, given one sample $X_i^A$, the sample $X_i^B$ forms a positive pair with it, and the other samples form negative pairs with $X_i^A$. The model will maximize the similarity of representations for the positive pair in contrast to the negative pairs.

## B. The Base Encoder

The base encoder takes the EEG signals $X_i^A$ or $X_i^B$ as inputs to generate aligned representations of the EEG data of different subjects. It adopts two one-dimensional convolutional operations to perform signal transformations (Fig. 1, upper-left). For simplicity, we omit the corner markers of $X_i^A$ or $X_i^B$ in the following sections.

### 1) Spatial convolution

As one EEG channel can pick up neural activities from multiple sources and the same source activity can influence signals of multiple EEG channels [65], we use a spatial convolution to transform the signals to a latent space for identifying plausible source activities. The spatial convolution is formulated as

$$H_{k:}^{(1)} = W_k X, k = 1, 2, \ldots, K_1 \qquad (1)$$

where $W_k \in \mathbb{R}^{1 \times M}$ is the weights of the $k$-th one-dimensional spatial convolution filter. There are $K_1$ filters in total. $H^{(1)} \in \mathbb{R}^{K_1 \times T}$ is the extracted representation by the spatial convolution. Each row of $H^{(1)}$ is a latent signal identified by a linear combination of the original signals $X \in \mathbb{R}^{M \times T}$. We denote the overall spatial convolution operation as $H^{(1)} = \text{Conv}_{\text{spat1}}(X)$ for simplicity.

### 2) Temporal convolution

Another property of the EEG signals is the dynamic changes over time [66], [67]. Therefore, we further apply a temporal convolution to learn the temporal patterns of the EEG signals:

$$H_{k_2, k_1, t} = v_{k_2} \cdot H^{(1)}_{k_1, t:t+P_1-1},$$

$$k_2 = 1, 2, \ldots, K_2; k_1 = 1, 2, \ldots, K_1; \ t = 1, 2, \ldots, T \qquad (2)$$

where $v_{k_2} \in \mathbb{R}^{P_1}$ represents the filter weights of the temporal convolution. The filter length is $P_1$, and the number of filters is $K_2$. $v_{k_2} \cdot H^{(1)}_{k_1, t:t+P_1-1}$ represents the dot product of the vectors $v_{k_2}$ and $H^{(1)}_{k_1, t:t+P_1-1}$. $H \in \mathbb{R}^{K_2 \times K_1 \times T}$ is the representation extracted by the temporal convolution. The input $H^{(1)}$ is padded to ensure the output $H$ is still of length $T$ on the temporal dimension. A temporal convolution filter can usually extract representations in a specific frequency band of the EEG signals. We denote the overall temporal convolution operation as $H = \text{Conv}_{\text{temp1}}(H^{(1)})$.

## C. The Projector

A nonlinear projector is utilized between the base encoder and the final contrastive loss (Fig. 1, upper-right). This idea is inspired by the SimCLR framework [55], in which a nonlinear projector can help the base encoder learn better representations for downstream prediction tasks. As the use of structure units is a common practice in neural network designs [68], [69], we employ a similar convolutional unit in the projector as that in the base encoder. Specifically, we adopt a spatial convolution to combine the amplitudes of different latent spatial components and a temporal convolution to extract the temporal patterns of amplitude change, with an average pooling layer prior to these convolution layers. The separable one-dimensional convolutions decouple the extraction of spatial patterns and temporal patterns with fewer parameters [70].

### 1) Average pooling

The average pooling has a kernel size of $1 \times S$ and a stride of $S$:

$$\hat{H}_{k_2, k_1, t} = mean\left(\delta\left(H_{k_2, k_1, [tS:(t+1)S]}\right)\right), k_2 = 1, 2, \ldots, K_2;$$

$$k_1 = 1, 2, \ldots, K_1; \ t = 1, 2, \ldots, \lfloor T/S \rfloor \qquad (3)$$

with the output $\hat{H} \in \mathbb{R}^{K_2 \times K_1 \times \lfloor T/S \rfloor}$. $\lfloor \cdot \rfloor$ means round down to the nearest integer. $\delta(\cdot)$ represents the exponential linear unit (ELU) [71].

### 2) Spatial convolution

Then a depthwise spatial convolution [72] is adopted to combine the pooled features across different latent signals:

$$G = \delta(\text{Conv}_{\text{spat2}}(\hat{H})). \qquad (4)$$

The function $\text{Conv}_{\text{spat2}}$ shares similar computations with $\text{Conv}_{\text{spat1}}$ but is a depthwise convolution with the filter size of $1 \times K_1$. Here, the input $\hat{H}$ is regarded as having $K_2$ feature maps, each of size $K_1 \times \lfloor T/S \rfloor$. In the depthwise convolution, each feature map of $\hat{H}$ is processed by $C$ spatial convolution filters, so there are $CK_2$ filters in total. $G \in \mathbb{R}^{CK_2 \times 1 \times \lfloor T/S \rfloor}$ is the output of the spatial convolution.

### 3) Temporal convolution

We further apply a depthwise temporal convolution to further identify the temporal patterns:

$$Z = \delta(\text{Conv}_{\text{temp2}}(G)) \qquad (5)$$

The function $\text{Conv}_{\text{temp2}}$ shares similar computations with $\text{Conv}_{\text{temp1}}$ but is a depthwise convolution with the filter size as $P_2$. There are $C$ temporal convolution filters for each feature map of $G$, resulting in $C \times CK_2 = C^2 K_2$ feature maps in the output $\boldsymbol{Z} \in \mathbb{R}^{C^2 K_2 \times 1 \times (\lceil T/S \rceil - P_2 + 1)}$. The depthwise convolutions reduce the parameter size and ensure specific spatiotemporal pattern extractions for each frequency band (extracted by temporal convolution in the base encuoder). Then, $\boldsymbol{Z}$ is transformed into a one-dimensional vector $\boldsymbol{z} \in \mathbb{R}^{C^2 K_2 (\lceil T/S \rceil - P_2 + 1)}$ for further similarity calculation.

### D. The Contrastive Loss

As introduced previously, the input samples $\mathcal{D} = \{\boldsymbol{X}_i^s | i = 1,2, \dots, N; s \in \{A, B\}\}$ are transformed into $\mathcal{D}_{latent} = \{\boldsymbol{z}_i^s | i = 1,2, \dots, N; s \in \{A, B\}\}$ via the base encoder and the projector. Then, the similarity of the input samples $\boldsymbol{X}_i^A$ and $\boldsymbol{X}_j^B$ is given by

$$sim(\boldsymbol{z}_i^A, \boldsymbol{z}_j^B) = \frac{\boldsymbol{z}_i^A \cdot \boldsymbol{z}_j^B}{\|\boldsymbol{z}_i^A\| \|\boldsymbol{z}_j^B\|}. \tag{6}$$

The contrastive loss aims to maximize the similarity of two pieces of EEG signals in a positive pair. Similar to the SimCLR framework [55], we adopt the normalized temperature-scaled cross-entropy loss computed by

$$l_i^A = -\log\left(\frac{exp(sim(\boldsymbol{z}_i^A, \boldsymbol{z}_i^B)/\tau)}{\sum_{j=1}^{N} \mathbb{1}_{[j \neq i]} exp(sim(\boldsymbol{z}_i^A, \boldsymbol{z}_j^A)/\tau) + \sum_{j=1}^{N} exp(sim(\boldsymbol{z}_i^A, \boldsymbol{z}_j^B)/\tau)}\right) \tag{7}$$

where $\mathbb{1}_{[j \neq i]} \in \{0, 1\}$ is an indicator function. It is set to 1 iff $j \neq i$. By minimizing the loss function, the model will increase the similarity between $\boldsymbol{z}_i^A$ and $\boldsymbol{z}_i^B$ in contrast to all other possible sample pairs involving $\boldsymbol{z}_i^A$. Finally, the total loss of the minibatch is

$$L = \sum_{i=1}^{N} l_i^A + \sum_{i=1}^{N} l_i^B \tag{8}$$

The overall training algorithm of the contrastive learning procedure is summarized in Algorithm 1.

### 3.2 The Prediction Procedure

In the prediction procedure, we use the trained base encoder to align the representations from different subjects and then extract predictive features for emotion recognition (Fig. 1).

Here, we denote the data in the prediction procedure as $\{\boldsymbol{X}^{pred}\}$ and their labels as $\{\boldsymbol{y}\}$. The label $\boldsymbol{y}$ is a categorical variable. For example, if there are three emotional categories, $\boldsymbol{y}$ can take three values: 0, 1, or 2. We need to predict the emotion category $\boldsymbol{y}$ for each sample $\boldsymbol{X}^{pred} \in \mathbb{R}^{M \times T'}$. CLISA extracts the aligned representations from the input $\boldsymbol{X}^{pred}$ by the trained base encoder:

$$\boldsymbol{H}^{pred} = Conv_{temp1}(Conv_{spat1}(\boldsymbol{X}^{pred})) \tag{9}$$

where $\boldsymbol{H}^{pred} \in \mathbb{R}^{K_2 \times K_1 \times T'}$ can be regarded as signals with $K_2$ frequency components, and each component has $K_1$ latent dimensions with a time length of $T'$. We assume that $\boldsymbol{H}^{pred}$ possesses a better representation than $\boldsymbol{X}^{pred}$ for cross-subject emotion recognition. Note that $T'$ can be different from the time length $T$ used in contrastive learning, as the convolution operators can receive data of various lengths.

Considering the limited amount of data in EEG emotion recognition, machine learning models tend to overfit the high-dimensional representations. To obtain low-dimensional relevant representations for emotion recognition, we extract the widely-used differential entropy (DE) features [9] from $\boldsymbol{H}^{pred}$. DE is defined as

$$\boldsymbol{H}_{K_2, k_1}^{DE} = \frac{1}{2}\log(2\pi e\sigma^2(\boldsymbol{H}_{k_2, k_1,:}^{pred})), \ k_1 = 1,2, \dots, K_1,$$
$$k_2 = 1,2, \dots, K_2 \tag{10}$$

where $\sigma^2$ is the variance of the signal and $\boldsymbol{H}^{DE} \in \mathbb{R}^{K_2 \times K_1}$. DE measures the complexity of the time series. It is equivalent to the logarithmic energy spectrum in a specific frequency band [10]. We call the DE features $\boldsymbol{H}^{DE}$ as "trained DE features" as it is extracted from the output of the trained base encoder. The trained DE features from consecutive samples within one trial of one subject are concatenated across time and smoothed with a linear dynamical system (LDS) model as in Zheng & Lu's work [14]. The smoothed features are reshaped into a one-dimensional vector $\boldsymbol{h}^{DE} \in \mathbb{R}^{K_2 K_1}$ as the classifier's input. Here, we use a three-layer multilayer perceptron (MLP) as the classifier. We denote the transformation of the classifier as

$$\boldsymbol{h}^{out} = F(\boldsymbol{h}^{DE}). \tag{11}$$

The MLP is optimized with a minibatch stochastic gradient descent algorithm and cross-entropy loss to learn the mapping from the DE features $\{\boldsymbol{h}^{DE}\}$ to the emotion labels $\{\boldsymbol{y}\}$. The algorithm of the prediction procedure is also summarized in Algorithm 1.

---

**Algorithm 1.** The Training Algorithm for CLISA

**The contrastive learning procedure**
**Inputs:** Training data $\{\boldsymbol{X}\}$, the learning rate $\alpha_1$, the batch size $2N$, the training epochs $T_1$.
1: Initialize parameters of the base encoder $\theta_B$ and the projector $\theta_P$.
2: **for** $epoch = 1$ to $T_1$ **do**
3:    **repeat**
4:      Sample two subjects $A, B$.
5:      Sample $2N$ EEG samples $\{\boldsymbol{X}_i^s | i = 1,2, \dots, N; s \in \{A, B\}\}$ from subjects $A, B$ with the data sampler.
6:      Obtain $\{\boldsymbol{z}_i^s | i = 1,2, \dots, N; s \in \{A, B\}\}$ by (1)-(5).
7:      Calculate loss $L$ by (6)-(8).
8:      $\theta_B \leftarrow \theta_B - \alpha_1 \frac{\partial L}{\partial \theta_B}, \ \theta_P \leftarrow \theta_P - \alpha_1 \frac{\partial L}{\partial \theta_P}$.
9:      $\alpha_1 \leftarrow F_\alpha(\alpha_1)$ according to cosine annealing scheme with warm restarts.
10:    **until** all possible pairs of subjects are enumerated.
11: **Outputs:** Parameters $\theta_B$.

**The prediction procedure** [a]
**Inputs:** Data $\{\boldsymbol{X}^{pred}\}$. Trained parameters of the base encoder $\theta_B$ and the classifier $\theta_M$.
1: Calculate $\boldsymbol{H}^{pred}$ by (9).
2: Extract DE features $\boldsymbol{H}^{DE}$ by (10).
3: Obtain smoothed DE features $\boldsymbol{h}^{DE}$ with linear dynamical systems.
4: Obtain predicted labels with the classifier by (11).

---

[a] Note that the classifier parameters $\theta_M$ were trained by the cross-entropy loss to output training labels $\{\boldsymbol{y}\}$ with inputs $\{\boldsymbol{h}^{DE}\}$ from the training data. We omit the training step in the algorithm for simplicity.

## 4 EXPERIMENTS

In our computational experiments, the proposed CLISA method was implemented and evaluated on two datasets: a new dataset THU-EP with 80 subjects [64] and the widely-used SEED dataset [9], [14]. The THU-EP dataset is expected to be a good benchmark for testing cross-subject emotion recognition models for its relatively larger number of subjects than most publicly available datasets.

This section will first introduce the THU-EP dataset in more detail and the SEED dataset in brief, then describe the data preprocessing procedure and the implementation details of our model. Finally, we will introduce the approach of performance evaluation, performance comparison, and spatiotemporal pattern analysis.

### 4.1 The THU-EP dataset

*Subjects.* Eighty college students (50 females, mean age = 20.16 years, ranging from 17 to 24 years) were recruited into the study. Informed consent was obtained from all subjects. The study was approved by the Ethics Committee of Tsinghua University.

*Stimuli.* In the experiments, 28 emotional video clips were used as the stimuli. There were 12 video clips for eliciting four negative emotions (i.e., anger, disgust, fear, and sadness, three clips for each category), 12 video clips for eliciting four positive emotions (i.e., amusement, joy, inspiration, and tenderness, three clips for each category), and four video clips with neutral emotion. Therefore, nine emotion categories were included in total. These emotional categories were expected to cover the daily experienced emotion to a large extent [73], [74], [75], [76]. The video clips were selected from the published emotional video datasets [73], [77], [78]. The duration of the videos is 67 seconds on average, ranging from 34 to 129 seconds.

*Experimental procedure.* Each subject watched the video clips in seven blocks. Each block contained four trials. The subjects watched one video clip for each trial and rated their emotional states afterward. The subject reported their emotional states on 12 emotion items (i.e., anger, disgust, fear, sadness, amusement, joy, inspiration, and tenderness, as well as arousal, valence, familiarity, and liking) on a scale of 0 to 7. The four video clips in one block had the same valence (i.e., positive, negative, or neutral) to avoid possible influence between consecutive video clips with different valence. The subjects were asked to solve 20 arithmetic problems between two blocks to prevent carry-over effects of valence across blocks [79].

*EEG recording.* EEG signals were recorded using a 32-channel wireless EEG system (NeuSen.W32, Neuracle, China) placed according to the international 10-20 system: Fp1/2, Fz, F3/4, F7/8, FC1/2, FC5/6, Cz, C3/4, T3/4, A1/2 (left and right mastoids), CP1/2, CP5/6, T5/6, Pz, P3/4, PO3/4, Oz, O1/2. The sampling rate was 250 Hz. The EEG signals were referenced to CPz with a forehead ground at AFz. Electrode impedances were kept below 10 kOhm for all electrodes throughout the experiment.

### 4.2 The SEED dataset

The SEED dataset is a widely used benchmark to evaluate emotion recognition algorithms [9], [14]. It contains the EEG data collected from 15 subjects (8 females, mean age = 23.27 years, std of age = 2.37 years). Each of them watched 15 film clips. These film clips elicited three kinds of emotion, including positive, neutral, and negative (five film clips for each emotion). Each film clip was selected to elicit a single desired target emotion. The duration of each film clip is 226 seconds on average, ranging from 185 seconds to 265 seconds. Each subject was required to carry out the experiments for three sessions. There was a one-week or longer time interval between two sessions. For each session, the subjects watched one film clip for each trial, resulting in 15 trials. EEG signals were recorded using an ESI NeuroScan System2 with 62 channels placed according to the international 10-20 system with a sampling rate of 1000 Hz.

### 4.3 Data Preprocessing

Data preprocessing was conducted with Fieldtrip [80] and NoiseTools [81] toolkits in Matlab.

For the THU-EP dataset, we first applied a bandpass filter from 0.05 to 47 Hz. Then independent component analysis (ICA) was applied to remove possible artifacts due to eye movements, muscle movements, or other environmental noise. We used the Infomax algorithm [82] in Fieldtrip for ICA, which minimizes the mutual information of different components with natural gradient approach [83]. The ICA algorithm received continuous data of one subject (without extracting data epochs) as inputs. Other settings were left as their defaults in Fieldtrip. A conservative criterion was applied to the ICA-based artifact rejection procedure: only the independent components (ICs) showing intense and persistent noise were removed. One or two ICs were removed per subject. After that, we implemented an automatic denoising procedure with NoiseTools to fix the data for spatially or temporally localized noises on a single-trial basis: We first interpolated the noisy channels by their three closest channels. If the proportion of outliers from one channel exceeded 30% of time points, the channel was defined as a noisy channel. The outliers were defined as those whose absolute values exceeded three times the median absolute value in the particular trial. After that, we fixed the remaining outliers with a threshold of 100 uV, which means if the difference of absolute values between two consecutive time points of one channel exceeded 100 uV, we replaced the value of the latter time point with that of the previous one. On average, 0.13 channels were interpolated, with a maximum of 3 channels per trial. 10.8% of all trials had at least one channel interpolated. The purpose of having the ICA-based and NoiseTools-based procedure was to keep (and fix) rather than reject the data, which was important for the present contrastive learning framework.

After the automatic denoising procedure, we applied a bandpass filter from 4 to 47 Hz and re-referenced the data to the common average. We used the last 30 seconds of each trial to ensure the elicited emotions were coherent and intense enough [23], [33], [73]. The EEG data from one subject was rejected due to strong power-line noise in all the EEG channels.

For the SEED dataset, the publicly available data have been downsampled to 200 Hz and filtered from 0 to 75 Hz by the data provider. The same denoising procedure was applied to the SEED dataset as the THU-EP dataset. Then we applied a bandpass filter from 4 to 47 Hz and re-referenced the data to the common average.

### 4.4 Implementation Details

In the base encoder for the THU-EP dataset: the filter size of the spatial convolution was set as the number of EEG channels ($M$=30); the temporal convolution filter length was set to 60, as a 60th-order finite impulse response filter is expected to provide the necessary support for extracting EEG signals in a specific frequency band; the number of spatial convolution filters ($K_1$) and temporal convolution filters ($K_2$) were both set to 16, which was expected to be sufficient to extract enough information for emotion-related neural representations.

In the projector for the THU-EP dataset: the spatial convolution filter size was set to 16 to match the spatial dimension of its input (i.e., the output of the base encoder, $K_1$=16); the temporal convolution filter length was set to 6, which was expected to extract the temporal patterns of the averaged features sufficiently; the kernel length of the average pooling (prior to the spatial and temporal filters) was empirically set to 30; the parameter $C$ that controls the number of spatial or temporal filters in the projector was empirically set to 2.

For the SEED dataset, most of the hyperparameters were set the same as for THU-EP, except that 1) the temporal filter lengths and the average pooling's kernel length were proportional to those for the THU-EP dataset according to their sampling frequencies (200 Hz for SEED and 250 Hz for THU-EP) and 2) the spatial filter size in the base encoder was set to match the number of EEG channels in SEED ($M$=62).

The hyperparameters of the implemented CLISA method for the two datasets are shown in Table 1. The temporal filter sizes in both the base encoder and the projector were further manipulated, and the models with the chosen parameters were able to obtain the top performances (Table S1, S2).

### TABLE 1
HYPERPARAMETER SETTINGS OF THE MODEL ARCHITECTURE

| | Hyperparameters | THU-EP | SEED |
|---|---|---|---|
| Base encoder | The spatial filter size $M$ | 30 | 62 |
| | The temporal filter size $P_1$ | 60 | 48 |
| | The number of spatial filters $K_1$ | 16 | 16 |
| | The number of temporal filters $K_2$ | 16 | 16 |
| Projector | Kernel length of average pooling $S$ | 30 | 24 |
| | The spatial filter size $K_1$ | 16 | 16 |
| | The temporal filter size $P_2$ | 6 | 4 |
| | The number of spatial filters $CK_2$ | 32 | 32 |
| | The number of temporal filters $C^2K_2$ | 64 | 64 |

The time length of the samples in contrastive learning was determined by a tradeoff between the training samples' number and adequate sample length. On the one hand, the sample needed to be long enough for the contrastive learning model to capture the underlying emotion effectively. On the other hand, the longer sample's length resulted in an insufficient number of training data. Based on this consideration, the time length of one sample was set to 5 seconds (with a time step of 2 seconds) for the THU-EP dataset and 30 seconds (with a time step of 15 seconds) for the SEED dataset in contrastive learning. As the length of one trial is 30 seconds in the THU-EP dataset, the sample number from one trial is $\lfloor(30\text{-}5)/2\rfloor$+1=13. The length of one trial ranges from 185 seconds to 265 seconds on the SEED dataset, so the sample number from one trial ranges from 11 to 16 (13 on average).

In the contrastive learning procedure, we used stratified normalization [84] during training. In stratified normalization, we concatenated the same channel of different samples from one subject in the minibatch together and conducted z-score normalization. The stratified normalization was applied to inputs of the base encoder, outputs of average pooling, and outputs of the temporal convolution in the projector. For optimization of the contrastive learning model, we trained the model for 100 epochs with early stopping (maximal tolerance of 30 epochs without validation accuracy increase). We used an Adam optimizer [85] with a cosine annealing learning rate scheduler and a three-time warm restart [86]. The initial learning rate was set to 0.0007, and the weight decay was set to 0.015 empirically.

In the prediction procedure, the input sample length was set to 1 second as in many previous studies [12], [14], [15]. For the extracted DE features, we conducted adaptive feature normalization, which adapted the mean and variance in z-score normalization online as new data come in [87], [88], [89]. Specifically, the initial mean and variance were defined as the training data's mean ($\bar{x}_{train}$) and variance ($s_{train}^2$). Then, with new data of the testing subject coming in, the mean and variance were updated as the weighted summation of $\bar{x}_{train}$ (or $s_{train}^2$) and current available testing data's mean $\bar{x}_{1:t}$ (or variance, $s_{1:t}^2$). The weights for $\bar{x}_{train}$ and $s_{train}^2$ decayed exponentially with the input of the testing data, with a decay rate $\eta$ set as 0.99 empirically.

For the MLP classifier in the prediction procedure, there were two hidden layers with 30 units for each. Rectified linear units (ReLUs) [90] were used between every two layers. We used cross-entropy loss and an Adam optimizer to optimize the parameters. The learning rate was set as 0.0005 empirically, and the weight decay was selected from 0.005, 0.011, 0.025, 0.056, and 0.125 by cross-validation. The batch size was set as 256 empirically. We trained the model for 100 epochs with early stopping.

### 4.5 Performance Evaluation

We implemented two tasks to evaluate the model. The first was the generic cross-subject emotion recognition task. In this task, we used 10-fold cross-subject cross-validation for the THU-EP dataset and leave-one-subject-out cross-validation for the SEED dataset. The other was a more challenging task, which we call the generalizability test. In this task, the model needed not only to be generalized to new subjects but also to be generalized to new stimuli. This task could test whether the model learned real subject-invariant emotional representations rather than overfitting the existing stimuli. In particular, we used 2/3 of the trials from the training subjects in training and used the other 1/3 of the trials from the testing subjects in testing. Thus, the stimuli in testing had never been accessed by the model in training. The training and testing subjects partition was the same as in the generic cross-subject emotion recognition task.

For the generic cross-subject emotion recognition task on the THU-EP dataset, we implemented the following two versions with respect to the number of emotion classes: 1) a basic version of binary classification for negative and positive emotional states and 2) a more challenging version of the nine-class emotional classification, including eight emotional classes mentioned above and neutral emotion. In the basic version, the samples from 12 trials that elicited anger, disgust, fear, and sadness were all labeled as negative emotions, while the samples from 12 trials eliciting amusement, joy, inspiration, and tenderness were labeled as positive emotions. The neutral emotion category was not included in the basic version due to an unbalanced number of trials (only four trials for eliciting neutral emotion). Regarding the SEED dataset, we implemented a three-class emotion classification for negative, neutral, and positive.

## 4.6 Performance Comparison

To investigate the effectiveness of our contrastive learning method, we compared it with several competing emotion recognition methods, namely differential entropy (DE) features with MLP classifier (denoted by DE+MLP), subspace alignment (SA) [52], correlated component analysis (CorrCA) [42], [91], and SeqCLR [60]. DE+MLP was a simple baseline with no inter-subject alignment. SA and CorrCA conducted inter-subject alignment other than contrastive learning. SeqCLR implemented an alternative contrastive learning strategy by data augmentation. Among them, SA [28], [51] and SeqCLR [60] have reached state-of-the-art performance in cross-subject emotion recognition.

In the DE+MLP baseline, we extracted DE features directly from preprocessed EEG data, normalized them adaptively, smoothed, and fed them into an MLP. The DE features were extracted from four frequency bands: theta (4-8 Hz), alpha (8-13 Hz), beta (13-30 Hz), and gamma (30-47 Hz). The hyperparameters of the classifier were the same as the proposed method.

Subspace alignment (SA) utilizes a linear transformation to align the source and target data in PCA subspace. For the implementation of SA, we projected the data of training subjects to be aligned with that of each testing subject in the PCA subspace. In the subspace, the spatial dimension of the data was reduced to 16, which was identical to the number of spatial convolution filters in our model. After the projection, we filtered the data into 16 frequency bands equally spaced between the data's frequency limits ([4 Hz, 47 Hz]). DE features were extracted from each frequency band. This process resulted in 256-dimension (16×16, the same as in our method) features. Then we normalized, smoothed the data, and submitted them to the classifier.

Correlated component analysis (CorrCA) maximizes the correlation of latent components extracted from multiple EEG records. This method has been utilized to reveal subject-invariant brain responses to emotion [91]. For the implementation of CorrCA, we identified the linear transformation of each subject that maximizes the ratio of between-subject to within-subject covariance. Similar to SA, we also retained 16 components and decomposed them into 16 frequency bands.

SeqCLR is a contrastive learning method for EEG classification [60]. It learned the similarity between augmented samples with the original ones and achieved state-of-the-art performance on the SEED dataset. We implemented five augmentation strategies Mohsenvand et al. [60] proposed, including amplitude scale, time shift, zero-masking, additive Gaussian noise, and band-stop filter. The parameters of augmentation were the same as the original paper. To make a fair comparison, we used the same architectures of the base encoder and the projector as our model here. The other hyperparameters and pipelines were the same as our method.

Paired t-tests were performed for the performance comparison between different methods. To account for multiple comparisons, we conducted Bonferroni correction and reported the corrected $p$-values.

## 4.7 Spatiotemporal pattern analysis

We used the integrated gradients [92] method to identify important features for the MLP classifier and examined the spatiotemporal characteristics of these features. The integrated gradients method accumulates the gradient of prediction outputs along the straight line from a reference input (with all entries as the minimum of the actual input) to the actual input. The importance index for each feature was defined as its corresponding accumulated gradients. To obtain the most powerful predictive capacity, we first identified the best training epoch for the contrastive learning procedure and the prediction procedure with cross-validation. Then we trained the models with those specific epochs using all data. The integrated gradient method was applied to the trained MLP classifier to derive the features with large importance indices for each emotion category.

For each important DE feature in MLP, we further identified its corresponding spatial and temporal filters in the base encoder. We obtained the spatial activation pattern $A_k$ from the spatial filter $W_k$ by: $A_k = \overline{\Sigma_X} W_k$ ( $k = 1,2,...,16$), where $\overline{\Sigma_X}$ is the mean covariance matrix of the EEG time series across subjects [93]. Then we source localized the spatial activation pattern with Brainstorm [94] and OpenMEEG [95] toolboxes in Matlab. The forward model was estimated by a three-layer (scalp, inner skull, outer skull) symmetric boundary element method based on ICBM152 brain template [96]. It generated 15002 fixed-orientation dipoles oriented normally to the cortex. A lead-field matrix linking the dipole activities to the EEG signals was obtained by the boundary element method. Then the linear inverse kernel that maps the EEG signals to the source activities was estimated by the sLORETA algorithm [97].

# 5 RESULTS

## 5.1 Emotion Recognition Performance on the THU-EP Dataset

Our model achieved a binary classification accuracy of 71.9±8.8% for discriminating the positive and negative emotional states on the THU-EP dataset (Table 2, Table S3). Compared with the control model (i.e., DE+MLP) that extracted DE features directly, CLISA presented a significant improvement of 6.7% ($t(78)$=4.19, Bonferroni corrected $p$<0.001). This comparison indicated that the representations learned by our contrastive learning method were more powerful than the simple DE features for cross-subject emotion recognition. CLISA also significantly outperformed other inter-subject alignment methods, including SA (65.5±9.7%, $t(78)$=5.01, corrected $p$<0.001), CorrCA (64.5±11.0%, $t(78)$=5.25, corrected $p$<0.001), and SeqCLR (64.0±9.8%, $t(78)$=6.41, corrected $p$<0.001), demonstrating the superiority of our contrastive learning strategy. The area under the receiver operating characteristic (ROC) curve further illustrated the effectiveness of the proposed method, as it presented noticeably better predictive power under all thresholds than the other competing baselines (Fig. 3).

### TABLE 2
THE BINARY CLASSIFICATION ACCURACIES OF DIFFERENT METHODS ON THE THU-EP DATASET

| Methods | Avg (%) | Std (%) |
|---|---|---|
| DE+MLP | 65.2 | 11.5 |
| SA | 65.5 | 9.7 |
| CorrCA | 64.5 | 11.0 |
| SeqCLR | 64.0 | 9.8 |
| CLISA (ours) | 71.9 | 8.8 |

To illustrate the effects of contrastive learning, we visualized the original features and pretrained features of five example subjects by t-sne embedding (Fig. 4). The original DE features extracted from EEG signals were scattered in t-sne embedded space separately for different subjects (Fig. 4a). In contrast, the trained DE features produced by CLISA were merged together, and different emotion categories remained separable (Fig. 4b), indicating that our model could alleviate subject discrepancy effectively without loss of emotional separability, thus facilitating the cross-subject emotion recognition.

The performance of the proposed CLISA method benefitted substantially from the increasing number of training subjects in contrastive learning (Fig. 5). To investigate the effects of training subjects' numbers, we randomly selected a subset of training subjects with different subject numbers (8, 16, 24, 32, 40, 48, 56, 64, or 72) in the contrastive learning procedure. In the prediction procedure, we used all training subjects, which ensured the difference is only induced by the number of subjects in contrastive learning. Besides, we also implemented a baseline with no contrastive learning, i.e., we randomly initialized the base encoder and extracted DE features from its output. The performances of CLISA under different numbers of training subjects were illustrated in Fig. 5. We observed that the performance of CLISA rises considerably with the increasing number of training subjects. Therefore, we could expect the CLISA method to learn better subject-invariant emotion representations with more subjects for contrastive learning.

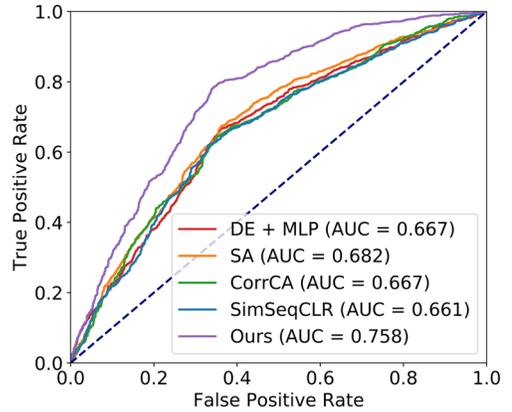

Fig. 3. The receiver operating characteristic (ROC) curve in the binary classification on the THU-EP dataset. The inset shows the area under the curve (AUC).

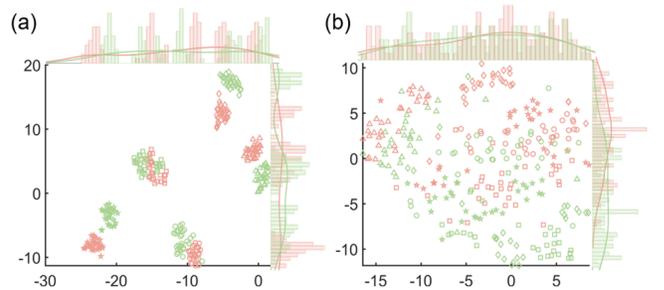

Fig. 4. (a) t-sne results of original differential entropy (DE) features. (b) t-sne results of DE features extracted from the output of trained base encoder (based on validation data of fold 1). Red: data points with negative emotion; Green: data points with positive emotion. Different marker types (circle, square, triangle, pentagram, and diamond) represent different subjects.

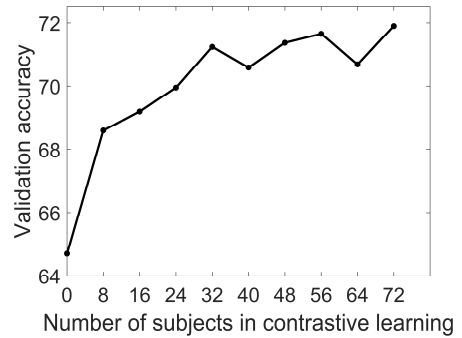

Fig. 5. The validation accuracy increases with the number of training subjects in contrastive learning.

In the generalizability test concerning new stimuli for the testing subjects (introduced in Section 4.5), our model achieved a classification accuracy of 63.4±17.1% (Table 3, Table S3), which was higher than all baseline models (the highest: 60.7±19.7%, although the improvement was not significant, corrected $ps>0.05$). This result indicated that the model did not just overfit or memorize the stimuli (i.e., the videos) that it had already seen in contrastive learning.

#### TABLE 3
THE GENERALIZABILITY TEST ON THE THU-EP DATASET

| Methods | Avg (%) | Std (%) |
|---|---|---|
| DE+MLP | 60.2 | 18.1 |
| SA | 60.7 | 19.7 |
| CorrCA | 59.0 | 17.3 |
| SeqCLR | 60.0 | 15.3 |
| CLISA (ours) | 63.4 | 17.1 |

*We used 16 trials of the training subjects and the other eight trials of the testing subjects in the generalizability test.*

In the nine-class emotional classification task, CLISA significantly surpassed all the other competing methods by an improvement of 10.2% (Table 4, Table S4, corrected $ps<0.001$ for all comparisons). The consistent improvement of CLISA over SA, CorrCA, and SeqCLR on these tasks shows the superiority of our method to other linear transformation methods or other contrastive learning strategies.

Based on the prediction results of CLISA, we further analyzed the confusion matrix of different emotion categories (Fig. 6). We observed that the disgust emotion could be identified with high accuracy, indicating its clear neural representation and high inter-subject consistency. Among positive emotions, amusement and tenderness could be identified best.

#### TABLE 4
THE NINE-CLASS CLASSIFICATION ACCURACIES OF DIFFERENT METHODS ON THE THU-EP DATASET

| Methods | Avg (%) | Std (%) |
|---|---|---|
| DE+MLP | 35.3 | 11.1 |
| SA | 35.5 | 11.8 |
| CorrCA | 34.5 | 10.4 |
| SeqCLR | 34.3 | 10.5 |
| CLISA (ours) | 45.7 | 11.8 |

### 5.2 Emotion Recognition Performance on the SEED Dataset

The effectiveness of our method was also evaluated on the widely-used SEED dataset. CLISA obtained an accuracy of 86.4±6.4% on the three-class emotion classification task (Table 5, Table S5). Similar to the THU-EP dataset, CLISA achieved better prediction performance than DE+MLP ($t(14)=3.37$, corrected $p=0.046$), SA ($t(14)=4.10$, corrected $p=0.011$), CorrCA ($t(14)=6.32$, corrected $p<0.001$), and SeqCLR ($t(14)=4.63$, corrected $p=0.004$). Furthermore, CLISA's performance was comparable with those of the latest models (i.e., DResNet and PPDA) reported in previous studies [26], [54]. It further demonstrated the effectiveness of our contrastive learning strategy compared to domain-adversarial strategies. The confusion matrix of our model is shown in Fig. 7. The model tended to confuse negative emotion with the other two emotions, especially with neutral emotion. The classification of the positive emotion was more accurate. These results were similar to previous studies [11], [13], [14].

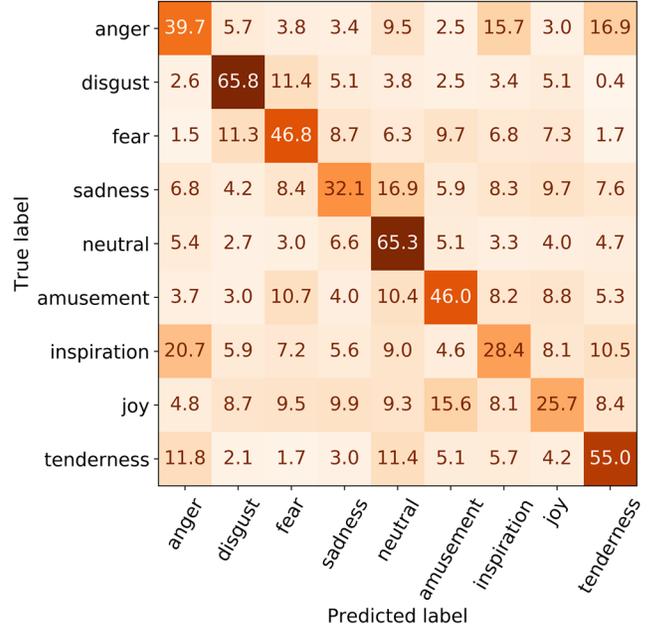

Fig. 6. The confusion matrix for the nine-class emotional classification of THU-EP dataset.

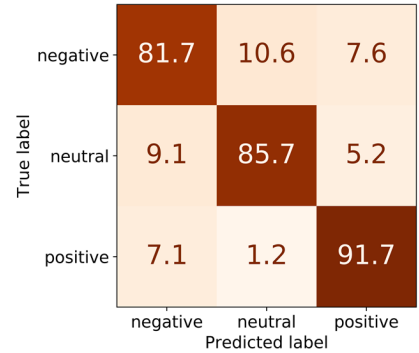

Fig. 7. The confusion matrix for the SEED dataset.

#### TABLE 5
THE CLASSIFICATION ACCURACIES OF DIFFERENT METHODS ON THE SEED DATASET

| Methods | Avg (%) | Std (%) |
|---|---|---|
| CLISA (ours) | 86.4 | 6.4 |
| DE+MLP | 79.9 | 8.7 |
| SA | 78.0 | 6.3 |
| CorrCA | 73.7 | 10.2 |
| SeqCLR | 78.4 | 9.2 |
| DResNet [26] | 85.3 | 8.0 |
| PPDA [54] | 86.7 | 7.1 |

*The comparison methods DE+MLP, SA, CorrCA, and SeqCLR were implemented in this paper. The results of DResNet and PPDA were reported in previous studies.*

In the generalizability test on the SEED dataset, our model achieved the highest classification accuracy as 77.4±13.4% (Table 6, Table S5), although the improvement was not statistically significant (corrected $ps>0.05$). This result further validated the generalizability of our model to new stimuli (i.e., the videos) that it had never seen.

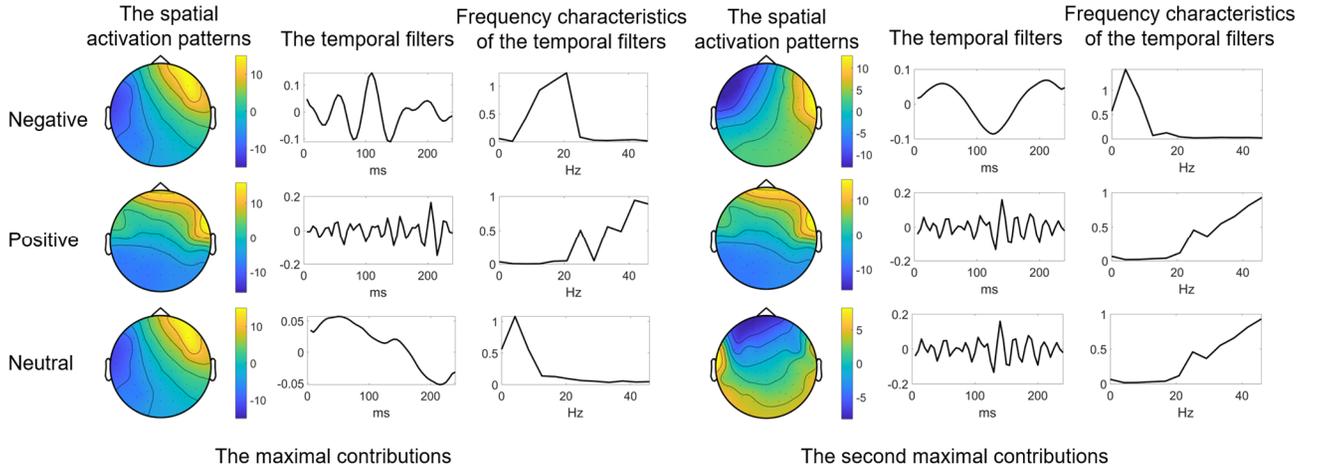

Fig. 8. Spatiotemporal characteristics of the important features for emotion classification on the SEED dataset. We visualize the spatial activation patterns, the temporal filters and the frequency characteristics of the temporal filters of the two most important features for negative, positive, and neutral emotions, respectively.

TABLE 6
THE GENERALIZABILITY TEST ON THE SEED DATASET

| Methods | Avg (%) | Std (%) |
|---|---|---|
| DE+MLP | 74.4 | 12.8 |
| SA | 73.9 | 8.9 |
| CorrCA | 70.9 | 13.6 |
| SeqCLR | 71.1 | 18.0 |
| CLISA (ours) | 77.4 | 13.4 |

*We used the first nine trials of the training subjects and the last six trials of the testing subjects in the generalizability test.*

## 5.3 Spatiotemporal Patterns for Emotion Recognition

This section presents the inter-subject aligned spatiotemporal representations extracted by CLISA. We analyzed the two most important features with the largest importance indices (See Section 4.7 for the definition) for each emotion category.

On the SEED dataset, we identified spatial activations mainly in anterior temporal regions for negative emotion, with frequency responses at around 12-21 Hz and 4 Hz (Fig. 8, Fig. S1). For positive emotion, the spatial activations focused on the temporal regions, with higher frequency responses of more than 25 Hz. The neutral state has frontal and temporal activations at around 4-8 Hz and left temporal activations at high frequencies of more than 25 Hz.

For binary classification on the THU-EP dataset, the spatial activations of the most important feature reside mainly in right occipital and right temporal regions, for both positive and negative emotions (Figs. S2, S3). The corresponding frequency response of this pattern was 4-8 Hz for positive emotion and less than 4 Hz for negative emotion. The second important features for positive and negative emotions also have similar spatial activation patterns located in bilateral temporal regions. The corresponding temporal response was around 21-38 Hz for positive emotion and 17-21 Hz for negative emotion. The activation in temporal regions with high-frequency responses for positive emotion was similar to that on the SEED dataset.

For nine-class emotion classification, the model identified distinct spatiotemporal patterns for different emotion categories. Only a few emotion pairs share exactly the same important spatiotemporal patterns (including disgust and fear, inspiration and neutral, and sadness and tenderness) (Figs. S4, S5). In comparison to spatiotemporal patterns in binary classification, anger has a similar pattern with the overall negative emotion pattern (occipital and posterior temporal activations with low-pass temporal filter). Joy has a similar pattern with the overall positive emotion pattern (occipital and posterior temporal activations with the frequency response at around 4-8 Hz). Other emotion categories have their own important spatiotemporal patterns, different from those for overall positive or negative emotions.

## 6 DISCUSSION

In comparison to existing cross-subject emotion recognition methods, our CLISA model has several prominent features. Firstly, the proposed contrastive learning strategy utilized the temporal alignment information of the data, i.e., which two pieces of data correspond to the same video segment. Thus, the contrastive learning strategy can match different subjects' data on a finer scale. As DResNet, PPDA, and other mainstream domain adaptation methods [38] were mostly based on domain classifiers, they can only roughly match the overall data distributions from different subjects. Secondly, CLISA implemented subject alignment with EEG time series as inputs, while DResNet, PPDA, and most existing cross-subject methods [38] used extracted features as inputs. To effectively process the EEG time series, CLISA used a CNN-based structure to filter the EEG time series and combine the spatial dimensions. In comparison, previous methods generally used MLP (e.g., DResNet) or LSTM (e.g., PPDA and BiDANN) for subject invariance learning and emotion classification. Our model demonstrated the possibility of finding a common representational space across subjects with EEG time series as inputs. Thirdly, CLISA can start recognizing a new user's emotion directly without accessing his/her data and improve the performance with more data coming in by adaptive feature normalization. In comparison, most domain adaptation methods require extensive data from the testing subjects for adaptation [38]. Therefore, CLISA can enhance the practicality of an emotion recognition system.

The spatial and temporal convolution architecture in CLISA offers the possibility to analyze activated sources and responsive frequencies for each emotion category. The spatial activations in temporal regions with high frequency responses (>20 Hz) were important for positive emotion on both the THU-EP and the SEED datasets. This result is in line with most previous studies [14], [15], [23]. The spatial activations for negative emotion were mainly in more anterior regions on the SEED dataset and more posterior regions on the THU-EP dataset. The discrepancy could be due to context-specific responses in each dataset that requires further investigations.

When we compare the important patterns for positive emotion and negative emotion, spatial activations generally have considerable overlap. It supports the affective workspace hypothesis that positive and negative emotion processing involves the same valence-general brain regions [98], which is consistent with a meta-analysis of 397 fMRI studies [99]. At the same time, the frequency responses for different affective states were distinct on both datasets. It indicates different dynamic processing mechanisms for positive and negative emotions, which fMRI studies can hardly reveal.

Furthermore, discrete emotion categories (in nine-class emotion classification) showed both emotion-specific spatial and temporal patterns. On the one hand, it suggests that from the perspective of EEG spatial activities, discrete emotion categories have more distinct neural representations than the affective dimension (i.e., valence). Different network activities underlying discrete emotion categories have also been well documented in fMRI studies [35], [100]. On the other hand, the temporal patterns are distinguishable for both discrete emotion categories and the valence dimension. Further studies could more thoroughly investigate the dynamic neural activities underlying emotion with EEG or MEG.

This study has some limitations that should be noted. Firstly, the proposed CLISA model was validated with EEG data of young adults (mean age = 20.16 years for the THU-EP dataset and mean age = 23.27 years for the SEED dataset). As age has been known to play an important role in emotion processing [101], [102], further studies are necessary to include subjects covering different age ranges for a more generalized model. Secondly, while the base encoder's architecture could be neurophysiologically meaningful, the possible neurophysiological implications for the projector were limited. Further studies are expected to develop a more neurophysiologically inspired projector, which could improve the interpretability of the whole network architecture. Finally, since CLISA aimed to differentiate different emotion categories, the spatiotemporal patterns that are shared across emotion categories might not be sufficiently revealed in this study. Further studies should be designed to identify both shared and distinguishable spatiotemporal patterns for emotion categories to better understand the neural mechanisms of emotion processing.

## 7 CONCLUSIONS

In this study, we proposed a contrastive learning method for inter-subject alignment, which is inspired by the inter-subject correlation studies in neuroscience. The proposed method achieved comparable or better performance in comparison to state-of-the-art methods on two datasets. On the THU-EP dataset, it obtained a binary classification accuracy of 71.9±8.8% and a nine-class classification accuracy of 45.7±11.8%. On the SEED dataset, it achieved a three-class classification accuracy of 86.4±6.4%. Moreover, we validated the model could generalize to unseen emotional stimuli better than other comparison methods. By visualizing important spatial and temporal filters in the model, we also demonstrated its potential for providing insights into the neural substrates of emotion.

## ACKNOWLEDGMENT

The authors wish to thank Wenyu Li, Xuefei Long, and Lilu Tang for EEG data collection, Weilong Zheng for providing the code of linear dynamical systems, Jin Liu, Xiaohe Li, and Likai Tang for their suggestions about the method.

This work was supported by the National Key Research and Development Program of China (Grant No. 2021ZD0200300), Spring Breeze Fund (2021Z99CFY037), National Natural Science Foundation of China (61977041, 61836004), Tsinghua University Initiative Scientific Research Program (20197010009), Center for Brain-inspired Computing Research, Beijing Innovation Center for Future Chips, Beijing Academy of Artificial Intelligence (BAAI), Guoqiang Institute, Tsinghua University, Key Scientific Technological Innovation Project by Ministry of Education, and the Open Project of Anhui Provincial Key Laboratory of Multimodal Cognitive Computation, Anhui University (MMC202001).

## REFERENCES

[1] L. Shu, J. Xie, M. Yang, Z. Li, Z. Li, D. Liao, X. Xu, and X. Yang, "A Review of Emotion Recognition Using Physiological Signals," Sensors, vol. 18, no. 7, p. 2074, 2018.

[2] S.M. Alarcao and M.J. Fonseca, "Emotions Recognition Using EEG Signals: A Survey," IEEE Trans. Affect. Comput., vol. 10, no. 3, pp. 374-393, 2017.

[3] C. Mühl, B. Allison, A. Nijholt, and G. Chanel, "A Survey of Affective Brain Computer Interfaces: Principles, State-of-the-Art, and Challenges," Brain-Comput. Interfaces, vol. 1, no. 2, pp. 66-84, 2014.

[4] Y. Ding, X. Hu, Z. Xia, Y.-J. Liu, and D. Zhang, "Inter-Brain EEG Feature Extraction and Analysis for Continuous Implicit Emotion Tagging During Video Watching," IEEE Trans. Affect. Comput., vol. 12, no. 1, pp. 92-102, 2018.

[5] X. Hu, J. Chen, F. Wang, and D. Zhang, "Ten Challenges for EEG-Based Affective Computing," Brain sci. adv., vol. 5, no. 1, pp. 1-20, 2019.

[6] R. Jenke, A. Peer, M. Buss, "Feature extraction and selection for emotion recognition from EEG. " IEEE Trans. Affect. Comput., vol. 5, no. 3, pp. 327-339, 2014.

[7] J. Cheng, M. Chen, C. Li, Y. Liu, R. Song, A. Liu, and X. Chen, "Emotion recognition from multi-channel eeg via deep forest," IEEE J. Biomed. Health Inform., vol. 15, no. 2, pp. 453-464, 2020.

[8] C. Li, Y. Hou, R. Song, J. Cheng, Y. Liu, and X. Chen, "Multi-channel EEG-based emotion recognition in the presence of noisy labels," Information Sciences, vol. 65, no. 140405, pp. 1-16, 2022.

[9] R-.N. Duan, J.-Y. Zhu, and B.-L. Lu, "Differential Entropy Feature for EEG-Based Emotion Classification," in Int. IEEE/EMBS Conf. Neural Eng. (NER), pp. 81-84, 2013.

[10] L.C. Shi, Y.Y. Jiao, and B.L. Lu, "Differential Entropy Feature for EEG-Based Vigilance Estimation," in Conf. Proc. IEEE Eng. Med. Biol. Soc. (EMBC), pp. 6627-6630, 2013.

[11] Y. Li, W. Zheng, Y. Zong, Z. Cui, T. Zhang, and X. Zhou, "A Bi-Hemisphere Domain Adversarial Neural Network Model for EEG Emotion Recognition," IEEE Trans. Affect. Comput., vol. 12, no. 2, pp. 494-504, 2018.

[12] Y. Li, L. Wang, W. Zheng, Y. Zong, L. Qi, Z. Cui, T. Zhang, and T. Song, "A Novel Bi-Hemispheric Discrepancy Model for EEG Emotion Recognition," IEEE Trans. Cogn. Develop., vol. 13, no. 2, pp. 354-367, 2021.

[13] P. Zhong, D. Wang, and C. Miao, "EEG-Based Emotion Recognition Using Regularized Graph Neural Networks," IEEE Trans. Affect. Comput., 2020, doi:10.1109/TAFFC.2020.2994159.

[14] W.-L. Zheng and B.-L. Lu, "Investigating Critical Frequency Bands and Channels for EEG-Based Emotion Recognition with Deep Neural Networks," IEEE Trans. Cogn. Develop. Syst., vol. 7, no. 3, pp. 162-175, 2015.

[15] W.-L. Zheng, J.-Y. Zhu, and B.-L. Lu, "Identifying Stable Patterns over Time for Emotion Recognition from EEG," IEEE Trans. Affect. Comput., vol. 10, no. 3, pp. 417-429, 2017.

[16] T. Song, W. Zheng, P. Song, and Z. Cui, "EEG Emotion Recognition Using Dynamical Graph Convolutional Neural Networks," IEEE Trans. Affect. Comput., vol. 11, no. 3, pp. 532-541, 2018.

[17] G. Zhang, M. Yu, Y.-J. Liu, G. Zhao, D. Zhang, and W. Zheng, "Sparsedgcnn: Recognizing Emotion from Multichannel EEG Signals," IEEE Trans. Affect. Comput., 2021, doi: 10.1109/TAFFC.2021.3051332.

[18] P. Li, H. Liu, Y. Si, C. Li, F. Li, X. Zhu, X. Huang, Y. Zeng, D. Yao, and Y. Zhang, "EEG Based Emotion Recognition by Combining Functional Connectivity Network and Local Activations," IEEE Trans. Biomed. Eng., vol. 66, no. 10, pp. 2869-2881, 2019.

[19] X. Shen, X. Hu, S. Liu, S. Song, and D. Zhang, "Exploring EEG Microstates for Affective Computing: Decoding Valence and Arousal Experiences During Video Watching," in Conf. Proc. IEEE Eng. Med. Biol. Soc. (EMBC), pp. 841-846, 2021.

[20] H. Cui, A. Liu, X. Zhang, X. Chen, K. Wang, and X. Chen, "EEG-based emotion recognition using an end-to-end regional-asymmetric convolution-al neural network," Knowl. Based. Syst., vol. 205, no. 106243, 2020.

[21] W. Tao, C. Li, R. Song, J. Cheng, Y. Liu, F. Wan, and X. Chen, "EEG-based emotion recognition via channel-wise attention and self attention," IEEE Trans. Affect. Comput., 2020, doi: 10.1109/TAFFC.2020.3025777.

[22] C. Li, Z. Zhang, R. Song, J. Cheng, Y. Liu, and X. Chen, "EEG-based Emotion Recognition via Neural Architecture Search," IEEE Trans. Affect. Comput., 2021, doi:10.1109/TAFFC.2021.3130387.

[23] S. Koelstra, C. Muhl, M. Soleymani, J. Lee, A. Yazdani, T. Ebrahimi, T. Pun, A. Nijholt, and I. Patras, "DEAP: A Database for Emotion Analysis using Physiological Signals," IEEE Trans. Affect. Comput., vol. 3, no. 1, pp. 18-31, 2012.

[24] S. Katsigiannis and N. Ramzan, "Dreamer: A Database for Emotion Recognition through EEG and ECG Signals from Wireless Low-Cost Off-the-Shelf Devices," IEEE J. Biomed. Health Inform., vol. 22, no. 1, pp. 98-107, 2017.

[25] J. Li, S. Qiu, Y.Y. Shen, C.L. Liu, and H. He, "Multisource Transfer Learning for Cross-Subject EEG Emotion Recognition," IEEE Trans. Cybern., vol. PP, no. 99, pp. 1-13, 2019.

[26] B.-Q. Ma, H. Li, W.-L. Zheng, and B.-L. Lu, "Reducing the Subject Variability of EEG Signals with Adversarial Domain Generalization," in Int. Conf. on Neural Inf. Process. (ICONIP), pp. 30-42, 2019.

[27] V. Jayaram, M. Alamgir, Y. Altun, B. Scholkopf, and M. Grosse-Wentrup, "Transfer Learning in Brain-Computer Interfaces," IEEE Comput. Intell. Mag., vol. 11, no. 1, pp. 20-31, 2016.

[28] J. Liu, X. Shen, S. Song, and D. Zhang, "Domain Adaptation for Cross-Subject Emotion Recognition by Subject Clustering," in Int. IEEE/EMBS Conf. Neural Eng. (NER), pp. 904-908, 2021.

[29] J. Li, S. Qiu, C. Du, Y. Wang, and H. He, "Domain Adaptation for EEG Emotion Recognition Based on Latent Representation Similarity," IEEE Trans. Cogn. Develop., vol. 12, no. 2, pp. 344-353, 2019.

[30] S. Hamann and T. Canli, "Individual Differences in Emotion Processing," Curr. Opin. Neurobiol., vol. 14, no. 2, pp. 233-238, 2004.

[31] G. Zhao, Y. Ge, B. Shen, X. Wei, and H. Wang, "Emotion Analysis for Personality Inference from EEG Signals," IEEE Trans. Affect. Comput., vol. 9, no. 3, pp. 362-371, 2017.

[32] W. Li, C. Wu, X. Hu, J. Chen, S. Fu, F. Wang, and D. Zhang, "Quantitative Personality Predictions from a Brief EEG Recording," IEEE Trans. Affect. Comput., 2020, doi: 10.1109/TAFFC.2020.3008775.

[33] W. Li, X. Hu, X. Long, L. Tang, J. Chen, F. Wang, and D. Zhang, "EEG Responses to Emotional Videos Can Quantitatively Predict Big-Five Personality Traits," Neurocomputing, vol. 415, pp. 368-381, 2020.

[34] R. Adolphs, "Neural Systems for Recognizing Emotion," Curr. Opin. Neurobiol., vol. 12, no. 2, pp. 169-177, 2002.

[35] P.A. Kragel and K.S. LaBar, "Decoding the Nature of Emotion in the Brain," Trends Cogn. Sci., vol. 20, no. 6, pp. 444-455, 2016.

[36] I.B. Mauss and M.D. Robinson, "Measures of Emotion: A Review," Cogn. Emot., vol. 23, no. 2, pp. 209-237, 2009.

[37] S. Hamann, "Mapping Discrete and Dimensional Emotions onto the Brain: Controversies and Consensus," Trends Cogn. Sci., vol. 16, no. 9, pp. 458-466, 2012.

[38] D. Wu, Y. Xu, and B.-L. Lu, "Transfer Learning for EEG-Based Brain-Computer Interfaces: A Review of Progress Made since 2016," IEEE Trans. Cogn. Develop., 2020.

[39] Y. Ganin, E. Ustinova, H. Ajakan, P. Germain, H. Larochelle, F. Laviolette, M. Marchand, and V. Lempitsky, "Domain-Adversarial Training of Neural Networks," J. Mach. Learn. Res., vol. 17, no. 1, pp. 2096-2030, 2016.

[40] D.A. Bridwell, C. Roth, C.N. Gupta, and V.D. Calhoun, "Cortical Response Similarities Predict Which Audiovisual Clips Individuals Viewed, but Are Unrelated to Clip Preference," PloS one, vol. 10, no. 6, p. e0128833, 2015.

[41] J.P. Dmochowski, M.A. Bezdek, B.P. Abelson, J.S. Johnson, E.H. Schumacher, and L.C. Parra, "Audience Preferences Are

Predicted by Temporal Reliability of Neural Processing," Nat. Commun., vol. 5, no. 1, pp. 1-9, 2014.

[42] J.P. Dmochowski, S. Paul, D. Joao, and L.C. Parra, "Correlated Components of Ongoing EEG Point to Emotionally Laden Attention – a Possible Marker of Engagement?," Front. Hum. Neurosci., vol. 6, 2012.

[43] U. Hasson, Y. Nir, I. Levy, G. Fuhrmann, and R. Malach, "Intersubject Synchronization of Cortical Activity During Natural Vision," Science, vol. 303, no. 5664, pp. 1634-1640, 2004.

[44] M. Kawasaki, Y. Yamada, Y. Ushiku, E. Miyauchi, and Y. Yamaguchi, "Inter-Brain Synchronization During Coordination of Speech Rhythm in Human-to-Human Social Interaction," Sci. Rep., vol. 3, no. 1, pp. 1-8, 2013.

[45] H. Liu, C. Zhao, F. Wang, and D. Zhang, "Inter-Brain Amplitude Correlation Differentiates Cooperation from Competition in a Motion-Sensing Sports Game," Soc. Cogn. Affect. Neurosci., vol. 16, no. 6, pp. 552-564, 2021.

[46] D. Zhang, "Computational EEG Analysis for Hyperscanning and Social Neuroscience," in Computational EEG Analysis: Springer, 2018, pp. 215-228.

[47] W.-L. Zheng and B.-L. Lu, "Personalizing EEG-Based Affective Models with Transfer Learning," in Int. Jt. Conf. Artif. Intell. (IJCAI), pp. 2732-2738, 2016.

[48] S.J. Pan, I.W. Tsang, J.T. Kwok, and Q. Yang, "Domain Adaptation Via Transfer Component Analysis," IEEE Trans. Neural Networks, vol. 22, no. 2, pp. 199-210, 2010.

[49] B. Scholkopf, A.J. Smola, and K. Muller, "Nonlinear Component Analysis as a Kernel Eigenvalue Problem," Neural Comput., vol. 10, no. 5, pp. 1299-1319, 1998.

[50] E. Sangineto, G. Zen, E. Ricci, and N. Sebe, "We Are Not All Equal: Personalizing Models for Facial Expression Analysis with Transductive Parameter Transfer," in Proc. ACM Int. Conf. Multimed. (MM), pp. 357-366, 2014.

[51] X. Chai, Q. Wang, Y. Zhao, Y. Li, D. Liu, X. Liu, and O. Bai, "A Fast, Efficient Domain Adaptation Technique for Cross-Domain Electroencephalography (EEG)-Based Emotion Recognition," Sensors, vol. 17, no. 5, p. 1014, 2017.

[52] B. Fernando, A. Habrard, M. Sebban, and T. Tuytelaars, "Unsupervised Visual Domain Adaptation Using Subspace Alignment," in IEEE Int. Conf. on Comput. Vis. (ICCV), pp. 2960-2967, 2014.

[53] X. Chai, Q. Wang, Y. Zhao, X. Liu, O. Bai, and Y. Li, "Unsupervised Domain Adaptation Techniques Based on Auto-Encoder for Non-Stationary EEG-Based Emotion Recognition," Comput. Biol. Med., vol. 79, pp. 205-214, 2016.

[54] L.-M. Zhao, X. Yan, and B.-L. Lu, "Plug-and-Play Domain Adaptation for Cross-Subject EEG-Based Emotion Recognition," in Proc. AAAI Conf. Artif. Intell. (AAAI), 2021.

[55] T. Chen, S. Kornblith, M. Norouzi, and G. Hinton, "A Simple Framework for Contrastive Learning of Visual Representations," Int. Conf. Mach. Learn. (ICML), pp. 1597-1607, 2020.

[56] J. Devlin, M.-W. Chang, K. Lee, and K. Toutanova, "Bert: Pre-Training of Deep Bidirectional Transformers for Language Understanding," in Proc. Conf. North American Chapter of the Association for Computational Linguistics (NAACL), pp. 4171-4186, 2019.

[57] P. Li, J. Wang, Y. Qiao, H. Chen, Y. Yu, X. Yao, P. Gao, G. Xie, and S. Song, "An Effective Self-Supervised Framework for Learning Expressive Molecular Global Representations to Drug Discovery," Brief. Bioinformatics, vol. 22, no. 6, 2021.

[58] X. Liu, Y. Luo, P. Li, S. Song, and J. Peng, "Deep Geometric Representations for Modeling Effects of Mutations on Protein-Protein Binding Affinity," PLoS Comput. Biol., vol. 17, no. 8, p. e1009284, 2021.

[59] X. Liu, F. Zhang, Z. Hou, L. Mian, Z. Wang, J. Zhang, and J. Tang, "Self-Supervised Learning: Generative or Contrastive," IEEE Trans. Knowl. Data Eng., 2021.

[60] M.N. Mohsenvand, M.R. Izadi, and P. Maes, "Contrastive Representation Learning for Electroencephalogram Classification," in Machine Learning for Health, pp. 238-253, 2020.

[61] H. Banville, O. Chehab, A. Hyvrinen, D.A. Engemann, and A. Gramfort, "Uncovering the Structure of Clinical EEG Signals with Self-Supervised Learning," J. Neural Eng., vol. 18, no. 4, p. 046020 (22pp), 2021.

[62] A. Oord, Y. Li, and O. Vinyals, "Representation Learning with Contrastive Predictive Coding," arXiv preprint arXiv:1807.03748, 2018.

[63] S.A. Nastase, V. Gazzola, U. Hasson, and C. Keysers, "Measuring Shared Responses across Subjects Using Intersubject Correlation," Soc. Cogn. Affect. Neurosci., vol. 14, no. 6, pp. 667-685, 2019.

[64] X. Hu, F. Wang, and D. Zhang, "Similar Brains Blend Emotion in Similar Ways: Neural Representations of Individual Difference in Emotion Profiles," NeuroImage, vol. 247, p. 118819, 2021.

[65] P.L. Nunez and R. Srinivasan, Electric Fields of the Brain: The Neurophysics of EEG. Oxford University Press, USA, 2006.

[66] T. König, F. Marti-Lopez, and P. Valdes-Sosa, "Topographic Time-Frequency Decomposition of the EEG," NeuroImage, vol. 14, no. 2, pp. 383-390, 2001.

[67] B. Crouch, L. Sommerlade, P. Veselcic, G. Riedel, B. Schelter, and B. Platt, "Detection of Time-, Frequency-and Direction-Resolved Communication within Brain Networks," Sci. Rep., vol. 8, no. 1, pp. 1-15, 2018.

[68] K. He, X. Zhang, S. Ren, and J. Sun, "Deep Residual Learning for Image Recognition," in Proc. IEEE Comput. Soc. Conf. Comput. Vis. Pattern Recognit. (CVPR), pp. 770-778, 2016.

[69] G. Huang, Z. Liu, L. Van Der Maaten, and K.Q. Weinberger, "Densely Connected Convolutional Networks," in Proc. IEEE Comput. Soc. Conf. Comput. Vis. Pattern Recognit. (CVPR), pp. 4700-4708, 2017.

[70] V.J. Lawhern, A.J. Solon, N.R. Waytowich, S.M. Gordon, C.P. Hung, and B.J. Lance, "EEGNet: A Compact Convolutional Network for EEG-Based Brain-Computer Interfaces," J. Neural Eng., vol. 15, no. 5, pp. 056013.1-056013.17, 2016.

[71] D.-A. Clevert, T. Unterthiner, and S. Hochreiter, "Fast and Accurate Deep Network Learning by Exponential Linear Units (Elus)," in Int. Conf. Learn. Represent. (ICLR), 2016.

[72] F. Chollet, "Xception: Deep Learning with Depthwise Separable Convolutions," in Proc. IEEE Comput. Soc. Conf. Comput. Vis. Pattern Recognit. (CVPR), pp. 1800-1807, 2017.

[73] X. Hu, J. Yu, M. Song, C. Yu, F. Wang, P. Sun, D. Wang, and D. Zhang, "EEG Correlates of Ten Positive Emotions," Front. Hum. Neurosci., vol. 11, p. 26, 2017.

[74] X. Hu, C. Zhuang, F. Wang, Y.-J. Liu, C.-H. Im, and D. Zhang, "Fnirs Evidence for Recognizably Different Positive Emotions," Front. Hum. Neurosci., vol. 13, p. 120, 2019.

[75] A. Mendiburo‐Seguel, D. Páez, and F. Martínez‐Sánchez, "Humor Styles and Personality: A Meta‐Analysis of the Relation between Humor Styles and the Big Five Personality Traits," Scand. J. Psychol., vol. 56, no. 3, pp. 335-340, 2015.

[76] P.E. Ekman and R.J. Davidson, The Nature of Emotion: Funda-mental Questions. Oxford University Press, 1994.

[77] A. Schaefer, F. Nils, X. Sanchez, and P. Philippot, "Assessing the Effectiveness of a Large Database of Emotion-Eliciting Films: A New Tool for Emotion Researchers," Cogn. Emot., vol. 24, no. 7, pp. 1153-1172, 2010.

[78] Y.-J. Liu, M. Yu, G. Zhao, J. Song, Y. Ge, and Y. Shi, "Real-Time Movie-Induced Discrete Emotion Recognition from EEG Sig-nals," IEEE Trans. Affect. Comput., vol. 9, no. 4, pp. 550-562, 2017.

[79] L.F. Van Dillen and S.L. Koole, "Clearing the Mind: A Working Memory Model of Distraction from Negative Mood," Emotion, vol. 7, no. 4, p. 715, 2007.

[80] R. Oostenveld, P. Fries, E. Maris, and J.-M. Schoffelen, "FieldTrip: Open Source Software for Advanced Analysis of MEG, EEG, and Invasive Electrophysiological Data," Comput. Intell. Neurosci., vol. 2011, no. 156869, 2011.

[81] A. de Cheveigné and D. Arzounian, "Robust Detrending, Reref-erencing, Outlier Detection, and Inpainting for Multichannel Data," NeuroImage, vol. 172, pp. 903-912, 2018.

[82] A.J. Bell, T.J. Sejnowski, "An information-maximization ap-proach to blind separation and blind deconvolution," Neural Comput., vol. 7, no. 6, pp. 1129-1159, 1995.

[83] S. Amari, A. Cichocki, H. Yang, "A new learning algorithm for blind signal separation," Adv. Neural Inf. Process. Syst. (Neu-rIPS), pp. 757-763, 1995.

[84] J. Fdez, N. Guttenberg, O. Witkowski, and A. Pasquali, "Cross-Subject EEG-Based Emotion Recognition through Neural Net-works with Stratified Normalization," Front. Neurosci., vol. 15, p. 11, 2021.

[85] D.P. Kingma and J. Ba, "Adam: A Method for Stochastic Optimi-zation," in Int. Conf. Learn. Represent. (ICLR), 2015.

[86] I. Loshchilov and F. Hutter, "Sgdr: Stochastic Gradient Descent with Warm Restarts," in Int. Conf. Learn. Represent. (ICLR), 2016.

[87] J. Bogaarts, D.M. Hilkman, E.D. Gommer, V. van Kranen-Mastenbroek, and J.P. Reulen, "Improved Epileptic Seizure De-tection Combining Dynamic Feature Normalization with EEG Novelty Detection," Med. Biol. Eng. Comput., vol. 54, no. 12, pp. 1883-1892, 2016.

[88] D. Bollegala, "Dynamic Feature Scaling for Online Learning of Binary Classifiers," Knowl. Based. Syst., vol. 129, pp. 97-105, 2017.

[89] E. Ogasawara, L.C. Martinez, D. De Oliveira, G. Zimbrão, G.L. Pappa, and M. Mattoso, "Adaptive Normalization: A Novel Data Normalization Approach for Non-Stationary Time Series," in Int. Jt. Conf. Neural Netw. (IJCNN), pp. 1-8, 2010.

[90] V. Nair and G.E. Hinton, "Rectified Linear Units Improve Re-stricted Boltzmann Machines," in Int. Conf. Mach. Learn. (ICML), pp. 807-814, 2010.

[91] L.C. Parra, "Correlated Components Analysis – Extracting Relia-ble Dimensions in Multivariate Data," arXiv preprint arXiv:1801.08881.

[92] M. Sundararajan, A. Taly, Q. Yan, "Axiomatic attribution for deep networks," Int. Conf. Mach. Learn. (ICML), pp. 3319-3328, 2017.

[93] S. Haufe, F. Meinecke, K. Görgen, S. Dähne, J.-D. Haynes, B. Blankertz, and F. Bießmann, "On the Interpretation of Weight Vectors of Linear Models in Multivariate Neuroimaging," Neu-roimage, vol. 87, pp. 96-110, 2014.

[94] F. Tadel, S. Baillet, J.C. Mosher, D. Pantazis, and R.M. Leahy, "Brainstorm: A User-Friendly Application for MEG/EEG Anal-ysis," Comput. Intell. Neurosci., vol. 2011, 2011.

[95] A. Gramfort, T. Papadopoulo, E. Olivi, and M. Clerc, "Open-MEEG: Opensource Software for Quasistatic Bioelectromagnet-ics," Biomed. Eng. Online, vol. 9, no. 1, pp. 1-20, 2010.

[96] G. Grabner, A.L. Janke, M.M. Budge, D. Smith, J. Pruessner, and D.L. Collins, "Symmetric Atlasing and Model Based Segmenta-tion: An Application to the Hippocampus in Older Adults," in Med. Image Comput. Comput. Assist. Interv. (MICCAI), pp. 58-66, 2006.

[97] R.D. Pascual-Marqui, "Standardized Low-Resolution Brain Elec-tromagnetic Tomography (Sloreta): Technical Details," Methods Find. Exp. Clin. Pharmacol., vol. 24, no. Suppl D, pp. 5-12, 2002.

[98] L.F. Barrett and E. Bliss-Moreau, "Affect as a Psychological Prim-itive," Adv. Exp. Soc. Psychol., vol. 41, pp. 167-218, 2009.

[99] K.A. Lindquist, A.B. Satpute, T.D. Wager, J. Weber, and L.F. Bar-rett, "The Brain Basis of Positive and Negative Affect: Evidence from a Meta-Analysis of the Human Neuroimaging Literature," Cereb. Cortex, vol. 26, no. 5, pp. 1910-1922, 2016.

[100] T.D. Wager, J. Kang, T.D. Johnson, T.E. Nichols, A.B. Satpute, and L.F. Barrett, "A Bayesian Model of Category-Specific Emo-tional Brain Responses," PLoS Comput. Biol., vol. 11, no. 4, p. e1004066, 2015.

[101] N.C. Ebner and H. Fischer, "Emotion and Aging: Evidence from Brain and Behavior," Front. Psychol., vol. 5, p. 996, 2014.

[102] V. Ferrari, N. Bruno, R. Chattat, and M. Codispoti, "Evaluative Ratings and Attention across the Life Span: Emotional Arousal and Gender," Cogn. Emot., vol. 31, no. 3, pp. 552-563, 2017.

## SUPPLEMENTARY MATERIALS

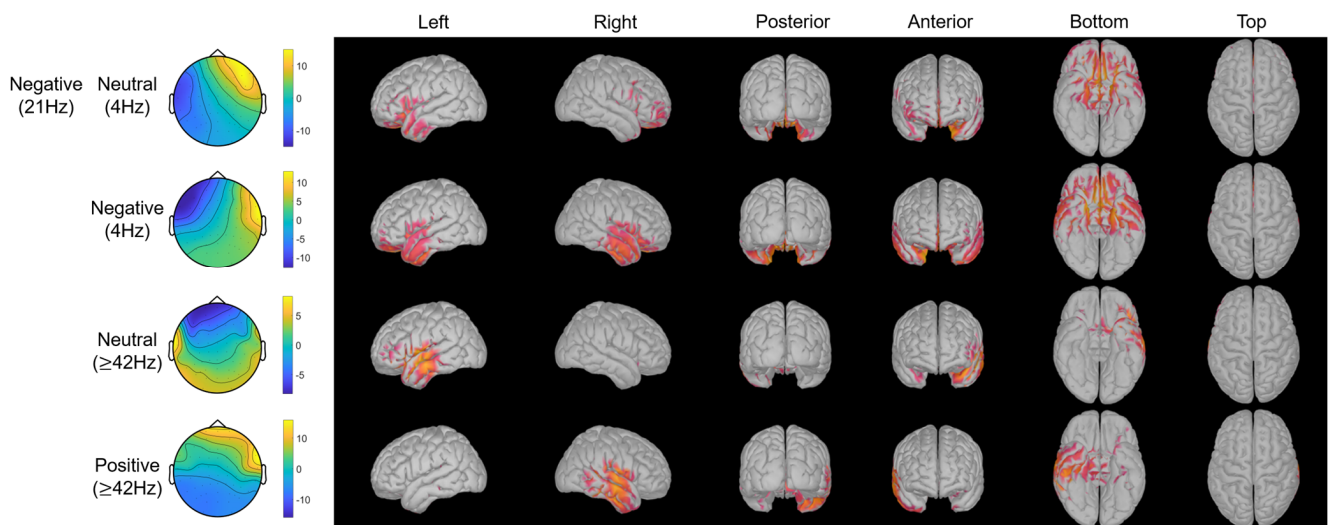

Figure S1. Source localization for the spatial activation patterns of the important features on the SEED dataset. The frequency below the emotion category (on the left) indicates the maximal frequency response of the corresponding temporal filter.

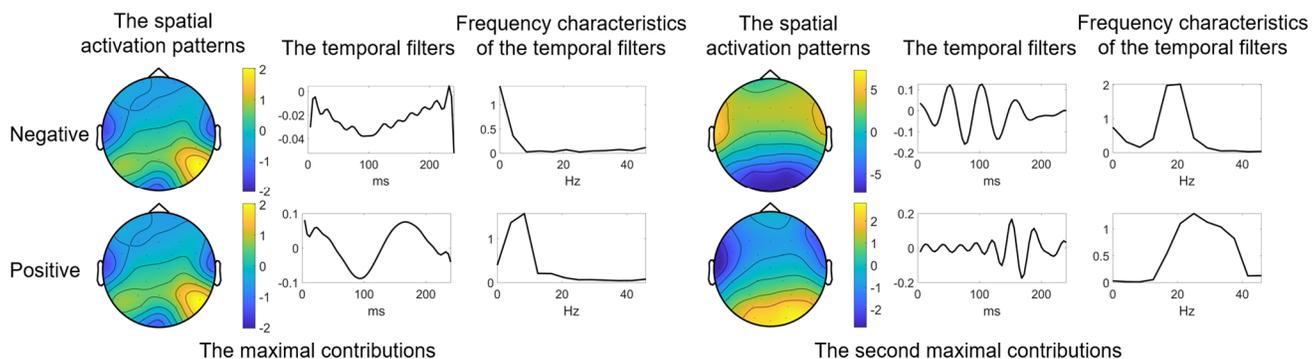

Figure S2. Spatiotemporal characteristics of the two most important features for binary emotion classification on the THU-EP dataset.

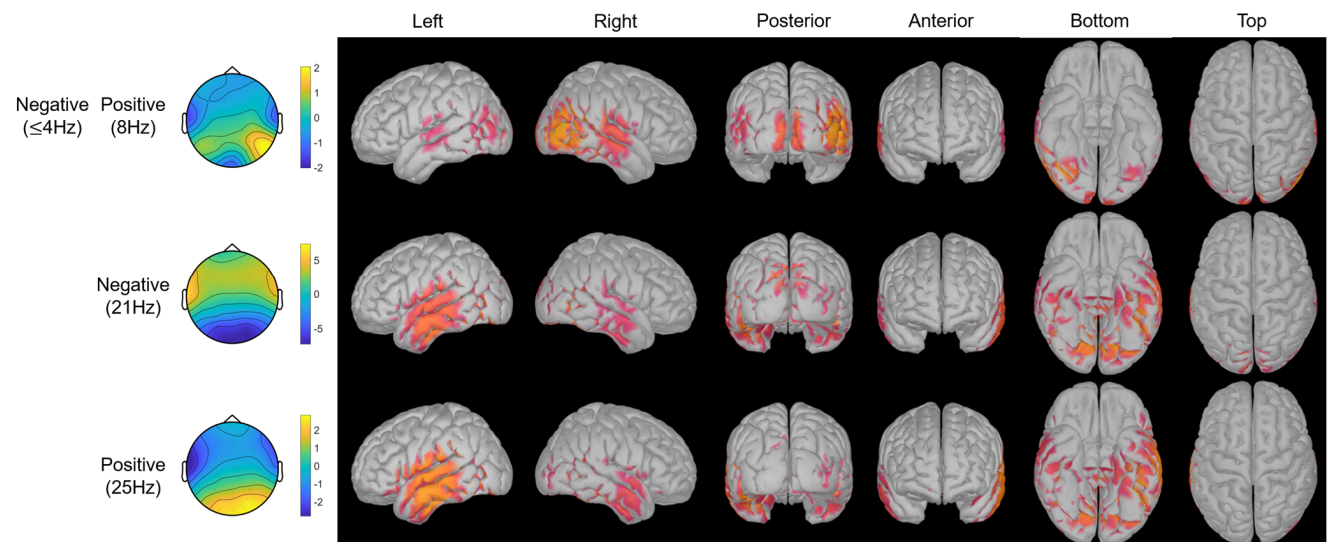

Figure S3. Source localization for the spatial activation patterns of the important features in binary emotion classification of the THU-EP dataset. The frequency below the emotion category (on the left) indicates the maximal frequency response of the corresponding temporal filter.

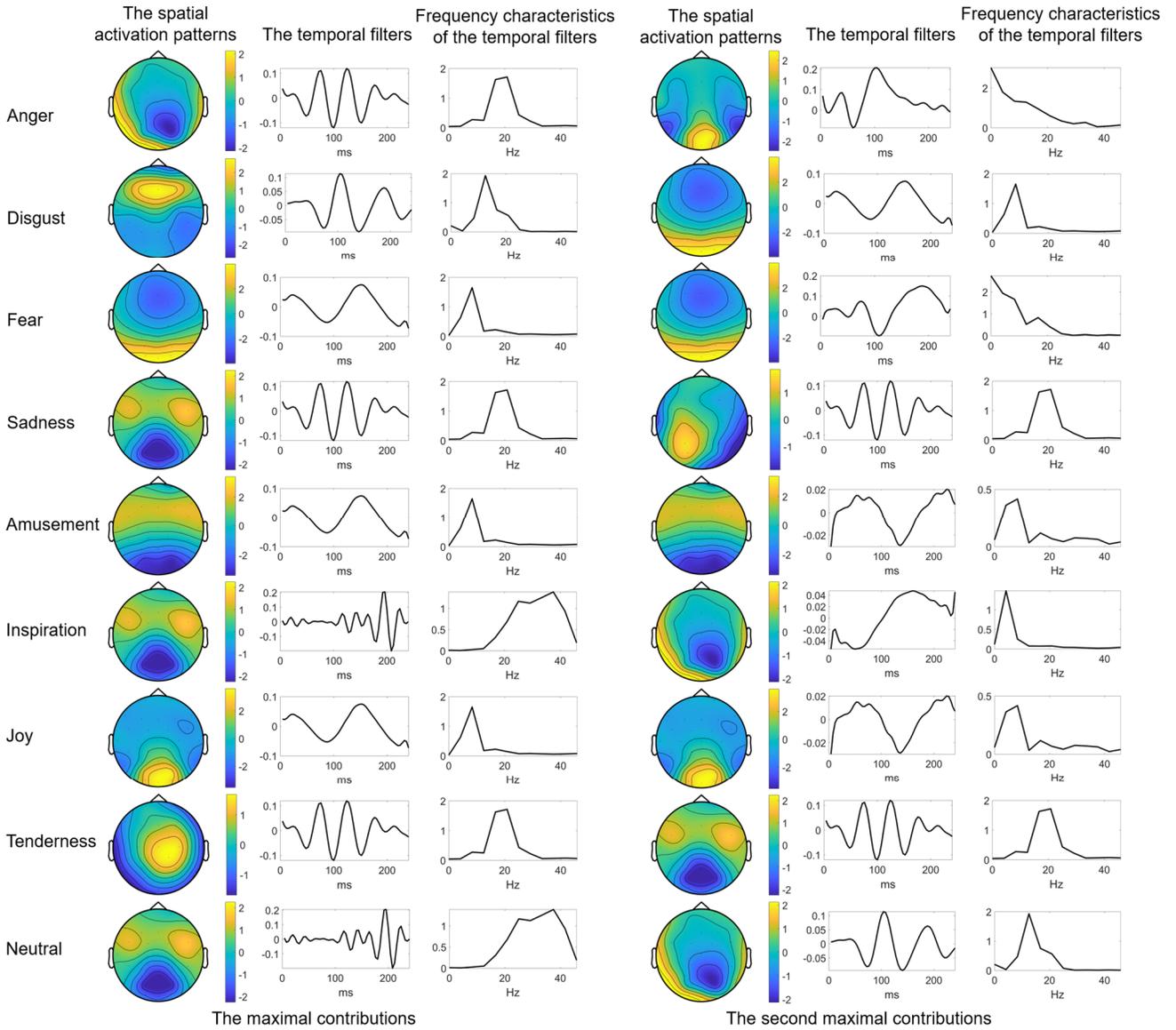

Figure S4. Spatiotemporal characteristics of the two most important features for nine-class emotion classification on the THU-EP dataset.

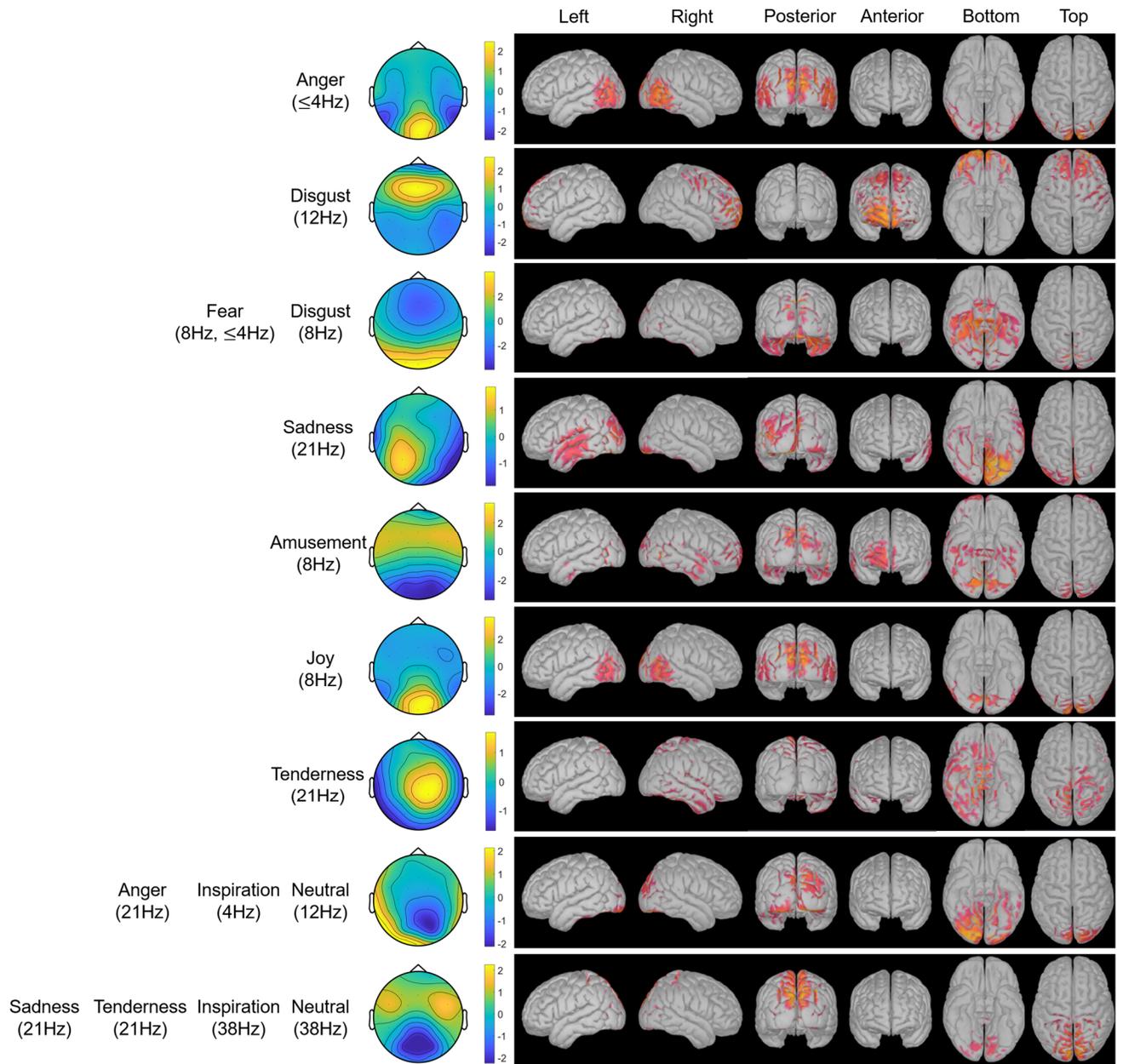

Figure S5. Source localization for the spatial activation patterns of the important features in nine-class emotion classification of the THU-EP dataset. The frequency below the emotion category (on the left) indicates the maximal frequency response of the corresponding temporal filter.

TABLE S1

THE PERFORMANCE WITH DIFFERENT TEMPORAL CONVOLUTION FILTER LENGTHS IN THE BASE ENCODER ON THE BINARY CLASSIFICATION FOR THE THU-EP DATASET

| Temporal Filter length in the base encoder | Avg (%) | Std (%) |
|---|---|---|
| 20 | 68.5 | 10.8 |
| 40 | 69.3 | 10.0 |
| 60 | 71.9 | 8.8 |
| 80 | 72.1 | 10.4 |
| 100 | 71.5 | 8.6 |

*The temporal filter length in the projector was set as 6 here.*

TABLE S2
THE PERFORMANCE WITH DIFFERENT TEMPORAL CONVOLUTION FILTER LENGTHS IN THE PROJECTOR ON THE BINARY CLASSIFICA-
TION FOR THE THU-EP DATASET

| Temporal Filter length in the projector | Avg (%) | Std (%) |
|---|---|---|
| 3 | 69.9 | 10.0 |
| 6 | 71.9 | 8.8 |
| 12 | 71.1 | 10.0 |
| 24 | 71.0 | 9.2 |

*The temporal filter length in the base encoder was set as 60 here.*

TABLE S3
THE BINARY CLASSIFICATION ACCURACIES FOR EACH SUBJECT IN THE THU-EP DATASET

| Subject No. | Binary Cross-subject emotion recognition | | | | | Generalizability test | | | | |
|---|---|---|---|---|---|---|---|---|---|---|
| | DE+MLP | SA | CorrCA | SeqCLR | CLISA | DE+MLP | SA | CorrCA | SeqCLR | CLISA |
| 1 | 66.7 | 67.8 | 62.5 | **70.8** | **70.8** | 62.5 | 50.0 | **75.0** | 62.5 | 62.5 |
| 2 | 75.0 | **84.4** | 75.0 | 83.3 | 78.1 | **75.0** | **75.0** | 50.0 | 62.5 | 62.5 |
| 3 | 62.5 | 70.0 | 54.2 | 62.5 | **83.3** | 25.0 | **62.5** | 37.5 | 37.5 | **62.5** |
| 4 | **75.0** | 54.4 | 62.5 | 66.7 | 70.8 | 70.8 | 50.0 | 75.0 | 62.5 | **100.0** |
| 5 | 45.8 | 67.9 | 58.3 | 58.3 | **83.3** | 75.0 | 62.5 | 62.5 | 62.5 | **87.5** |
| 6 | 79.2 | 77.8 | **87.5** | 66.7 | 66.7 | **87.5** | 62.5 | 75.0 | 75.0 | 75.0 |
| 7 | **75.0** | 53.4 | 62.5 | 73.5 | 70.8 | 62.5 | 62.5 | **62.5** | 49.6 | 62.5 |
| 8 | 62.5 | 61.3 | 58.3 | 58.3 | **79.2** | 62.5 | 75.0 | **100.0** | 62.5 | 50.0 |
| 9 | 54.2 | 68.8 | 74.6 | 62.5 | **75.0** | 62.5 | 62.5 | 50.0 | **75.0** | **75.0** |
| 10 | **75.0** | 66.1 | 55.1 | 56.0 | 66.7 | 37.5 | 37.5 | 37.5 | **62.5** | 37.5 |
| 11 | **75.0** | 70.0 | 66.7 | 62.5 | 70.8 | 62.5 | **75.0** | 62.5 | **75.0** | 37.5 |
| 12 | 54.2 | 65.2 | 66.7 | 66.7 | **70.8** | 87.5 | **100.0** | 50.0 | 75.0 | 75.0 |
| 13 | 62.5 | **67.9** | 61.9 | 62.5 | 66.7 | 50.0 | 62.5 | 50.0 | 62.5 | **87.5** |
| 14 | 79.2 | 72.7 | **79.2** | 75.0 | 75.0 | 62.5 | 50.0 | **75.0** | **75.0** | 50.0 |
| 15 | 62.5 | 46.9 | 62.5 | **70.8** | **70.8** | 62.5 | 12.5 | **75.0** | 62.5 | 50.0 |
| 16 | 83.3 | **85.6** | 62.5 | 79.2 | 83.3 | 75.0 | 75.0 | 50.0 | 50.0 | **87.5** |
| 17 | 62.5 | 60.1 | 62.5 | 58.3 | **70.8** | **75.0** | 25.0 | 50.0 | **75.0** | **75.0** |
| 18 | 58.3 | **84.4** | 75.0 | 58.3 | 71.8 | 62.5 | **75.0** | 37.5 | **75.0** | 50.0 |
| 19 | **75.0** | 54.4 | 50.0 | 45.8 | 62.5 | **75.0** | 50.0 | 50.0 | 62.5 | 50.0 |
| 20 | 70.8 | 61.2 | 79.2 | 83.3 | **87.5** | 50.0 | 50.0 | 62.5 | 62.5 | **75.0** |
| 21 | 41.7 | 51.3 | 45.8 | 58.3 | **66.7** | 75.0 | **87.5** | 75.0 | 62.5 | **87.5** |
| 22 | 54.2 | 62.9 | 70.8 | 66.7 | **79.2** | 37.5 | 37.5 | **75.0** | 37.5 | **75.0** |
| 23 | 47.4 | 66.1 | 58.3 | 67.4 | **75.0** | 62.5 | **87.5** | 37.5 | 50.0 | **87.5** |
| 24 | 70.8 | 81.7 | 79.2 | 75.0 | 79.2 | **100.0** | 62.5 | 75.0 | 75.0 | **100.0** |
| 25 | 58.3 | 67.0 | **75.0** | 58.3 | 70.8 | 25.0 | **50.0** | **50.0** | 37.5 | **50.0** |
| 26 | 75.0 | 70.9 | 62.5 | **79.2** | 75.0 | 62.5 | 62.5 | 62.5 | 62.5 | **75.0** |
| 27 | 62.5 | 70.9 | 70.8 | **75.0** | 66.7 | 62.5 | 75.0 | 37.5 | 50.0 | **87.5** |
| 28 | 54.2 | **67.8** | 66.7 | 58.3 | 66.7 | 62.5 | 62.5 | 62.5 | **75.0** | **75.0** |
| 29 | 62.5 | 63.1 | 79.2 | 50.0 | **91.7** | 37.5 | 37.5 | 62.5 | 37.5 | **75.0** |
| 30 | 83.3 | **83.5** | 54.2 | 58.3 | 70.8 | 62.5 | 50.0 | 62.5 | **87.5** | 50.0 |
| 31 | 70.8 | 73.9 | **83.3** | 50.0 | 75.0 | 37.5 | 37.5 | 25.0 | **50.0** | **50.0** |
| 32 | 54.2 | 58.2 | **70.8** | 62.5 | 66.7 | 50.0 | 62.5 | **75.0** | 54.6 | 56.7 |
| 33 | 54.2 | 55.2 | 58.3 | 42.5 | **75.0** | **75.0** | 62.5 | 50.0 | 62.5 | 50.0 |
| 34 | 62.5 | 52.2 | **66.7** | 54.2 | 50.0 | 37.5 | **50.0** | **50.0** | **50.0** | 25.0 |
| 35 | 86.1 | 58.3 | 70.8 | 79.2 | **86.4** | 75.0 | 62.5 | 62.5 | 62.5 | 62.5 |
| 36 | 54.2 | 70.9 | 41.7 | 58.3 | **71.1** | **50.0** | 37.5 | **50.0** | **50.0** | **50.0** |
| 37 | 47.8 | **73.9** | 54.2 | 50.1 | 66.7 | 65.8 | 62.5 | 62.5 | 62.5 | 62.5 |
| 38 | 66.7 | 69.3 | 83.3 | 58.3 | **87.5** | 75.0 | 75.0 | **87.5** | 62.5 | **87.5** |
| 39 | 66.7 | **75.7** | 58.3 | 62.5 | 70.0 | 37.5 | 37.5 | **62.5** | 37.5 | 25.0 |
| 40 | 63.2 | 50.8 | 50.0 | **75.0** | 66.7 | **75.0** | **75.0** | 62.5 | **75.0** | 62.5 |
| 41 | 58.3 | 66.0 | 70.8 | 75.0 | **79.2** | 12.5 | **100.0** | 37.5 | 37.5 | 87.5 |
| 42 | **83.3** | 73.9 | 50.0 | 75.0 | 70.8 | **100.0** | 62.5 | 62.5 | 75.0 | 50.0 |
| 43 | **83.3** | 73.9 | 66.7 | 75.0 | 79.2 | 62.5 | **75.0** | 37.5 | 50.0 | 50.0 |

| | | | | | | | | | |
|---|---|---|---|---|---|---|---|---|---|
| 44 | 73.2 | 74.8 | 57.8 | 66.7 | **83.3** | **93.3** | 75.0 | 87.5 | 75.0 | 75.0 |
| 45 | 58.3 | **75.7** | 62.5 | 70.8 | 70.8 | 50.0 | 75.0 | **87.5** | 62.5 | 75.0 |
| 46 | 45.0 | 49.6 | 50.0 | 50.0 | **66.7** | 37.5 | **75.0** | 50.0 | 50.0 | 62.5 |
| 47 | 54.2 | 63.1 | **75.0** | 66.7 | **75.0** | 50.0 | **68.8** | 64.6 | 62.5 | 62.5 |
| 48 | **91.7** | 50.5 | 66.7 | 53.5 | 75.0 | **87.5** | 12.5 | **87.5** | 75.0 | 62.5 |
| 49 | 75.0 | 73.9 | 75.0 | 75.0 | **79.2** | 50.0 | 50.0 | **62.5** | 37.5 | 25.0 |
| 50 | **70.8** | 66.9 | **70.8** | 58.3 | **70.8** | 75.0 | **87.5** | 75.0 | 75.0 | 62.5 |
| 51 | 33.3 | 46.6 | 54.2 | 62.2 | **75.0** | 37.5 | 50.0 | 25.0 | 62.5 | **75.0** |
| 52 | **69.9** | 58.2 | 65.1 | 45.8 | 62.5 | 75.0 | 50.0 | **100.0** | 62.5 | 50.0 |
| 53 | 58.3 | 52.2 | 62.5 | **70.8** | **70.8** | 75.0 | 50.0 | 75.0 | **87.5** | 62.5 |
| 54 | 41.7 | 60.1 | 66.7 | 58.3 | **79.2** | 62.5 | 37.5 | 50.0 | 50.0 | **75.0** |
| 55 | 75.0 | 71.8 | **87.5** | 66.7 | 70.8 | **75.0** | 37.5 | 62.5 | 25.0 | 37.5 |
| 56 | 66.7 | 66.1 | **70.8** | 58.3 | **70.8** | 37.5 | 62.5 | 62.5 | 58.8 | **75.0** |
| 57 | 66.7 | 64.0 | **70.8** | 66.7 | **70.8** | 37.5 | 25.0 | **62.5** | 50.0 | 62.5 |
| 58 | 58.3 | 48.7 | 45.8 | **73.1** | 66.7 | 62.5 | 50.0 | 25.0 | **64.6** | 37.5 |
| 59 | 66.7 | **70.9** | 70.8 | 45.8 | 66.7 | **62.5** | 60.4 | 37.5 | 50.0 | 50.0 |
| 60 | 70.8 | 67.8 | **79.2** | 70.8 | 47.4 | 62.5 | 62.5 | **75.0** | **75.0** | 62.5 |
| 61 | 62.5 | 60.0 | 62.5 | **70.8** | **70.8** | 75.0 | 75.0 | 37.5 | 50.0 | **87.5** |
| 62 | **62.5** | 59.2 | 45.8 | 41.7 | 50.0 | 37.5 | **87.5** | 50.0 | 37.5 | 62.5 |
| 63 | 66.7 | 66.0 | **75.0** | 70.8 | **75.0** | 37.5 | 37.5 | **62.5** | 37.5 | 50.0 |
| 64 | 68.2 | **84.4** | 58.3 | 66.7 | 79.2 | 62.5 | 75.0 | 50.0 | **87.5** | 62.5 |
| 65 | **70.8** | 69.1 | **70.8** | 62.2 | **70.8** | 62.5 | 62.5 | 50.0 | 62.5 | **75.0** |
| 66 | 43.9 | **75.7** | 75.0 | 75.0 | 70.8 | 75.0 | **87.5** | 25.0 | 62.5 | 50.0 |
| 67 | 62.5 | 58.3 | 63.6 | 50.0 | **66.7** | 37.5 | 50.0 | **62.5** | 50.0 | 62.5 |
| 68 | 58.3 | 65.8 | 54.2 | 75.0 | **91.7** | 75.0 | 75.0 | 55.4 | **87.5** | **87.5** |
| 69 | 70.8 | 70.8 | **75.0** | 70.8 | 70.8 | **75.0** | 50.0 | 62.5 | 62.5 | 62.5 |
| 70 | **79.2** | 75.7 | 62.5 | 66.7 | 68.9 | 37.5 | **62.5** | 50.0 | 50.0 | 37.5 |
| 71 | 54.2 | 65.4 | **70.8** | 66.7 | **70.8** | 62.5 | 75.0 | **82.1** | 75.0 | 50.0 |
| 72 | **79.2** | 58.3 | 41.7 | 75.0 | 62.5 | 50.0 | 50.0 | 50.0 | 25.0 | **75.0** |
| 73 | **66.7** | 55.4 | 41.7 | 45.8 | 46.3 | 62.5 | **87.5** | **87.5** | 75.0 | 75.0 |
| 74 | 62.5 | 56.1 | **70.8** | 66.7 | 62.5 | 75.0 | **87.5** | 50.0 | 62.5 | 62.5 |
| 75 | **70.8** | 69.1 | 53.1 | **70.8** | 66.1 | **62.5** | 37.5 | 37.5 | 25.0 | 37.5 |
| 76 | 79.2 | **79.6** | 75.0 | 58.3 | 70.8 | 75.0 | **100.0** | 87.5 | 75.0 | 87.5 |
| 77 | 54.2 | 48.3 | **58.3** | **58.3** | 54.2 | 37.5 | 12.5 | **75.0** | 62.5 | 62.5 |
| 78 | 75.0 | 54.3 | 41.7 | 67.2 | **83.3** | 75.0 | 75.0 | 50.0 | **87.5** | 50.0 |
| 79 | 75.0 | 66.1 | 66.7 | 54.2 | **83.3** | 25.0 | **87.5** | 37.5 | 37.5 | 62.5 |
| Avg | 65.2 | 65.5 | 64.5 | 64.0 | **71.9** | 60.2 | 60.7 | 59.0 | 60.0 | **63.4** |
| Std | 11.5 | 9.7 | 11.0 | 9.8 | 8.8 | 18.1 | 19.7 | 17.3 | 15.3 | 17.1 |
| $t(78)$ | 4.19 | 5.01 | 5.25 | 6.41 | - | 1.28 | 1.12 | 1.70 | 1.43 | - |
| Bonferroni corrected $p$ | <0.001 | <0.001 | <0.001 | <0.001 | - | 1.000 (0.206) [a] | 1.000 (0.268) | 0.940 | 1.000 (0.158) | - |

We reported t-values and Bonferroni corrected p-values in the paired t-tests between the performance of CLISA and other models.
[a] The p-values in parentheses were uncorrected.

TABLE S4
THE NINE-CLASS CLASSIFICATION ACCURACIES FOR EACH SUBJECT IN THE THU-EP DATASET

| Subject No. | DE+MLP | SA | CorrCA | SeqCLR | CLISA |
|---|---|---|---|---|---|
| 1 | 28.6 | 23.4 | 28.6 | 37.4 | **42.9** |
| 2 | 21.4 | 45.9 | 28.6 | 25.0 | **46.4** |
| 3 | **53.2** | 32.2 | 28.6 | 46.4 | 35.7 |
| 4 | 35.7 | 42.7 | **42.9** | 24.4 | 28.6 |
| 5 | 35.1 | 31.5 | 17.9 | 28.6 | **57.1** |
| 6 | 28.6 | 22.1 | **46.4** | 35.7 | 36.7 |
| 7 | **50.0** | 29.4 | 32.1 | 35.2 | **50.0** |
| 8 | 50.0 | 36.8 | 50.0 | 48.2 | **60.7** |
| 9 | 32.1 | 59.0 | 42.9 | 32.1 | **60.7** |
| 10 | 21.4 | 11.7 | 17.9 | 25.0 | **46.4** |
| 11 | 45.5 | 50.2 | 39.3 | 32.1 | **60.7** |
| 12 | 28.6 | 32.2 | 32.1 | 27.0 | **50.0** |
| 13 | 50.0 | 41.3 | 35.0 | 32.1 | **52.4** |
| 14 | 50.0 | **64.8** | 42.9 | 50.0 | 60.2 |
| 15 | 39.3 | 35.5 | 39.3 | 39.3 | **57.1** |
| 16 | 28.6 | 23.4 | 44.6 | 42.1 | **46.4** |
| 17 | 39.3 | 47.2 | 35.7 | 37.5 | **57.1** |
| 18 | **50.0** | 22.8 | 32.1 | 28.6 | 47.4 |
| 19 | 32.9 | 44.3 | 39.3 | 46.4 | **64.3** |
| 20 | 46.4 | 39.7 | 54.5 | 53.6 | **67.9** |
| 21 | 21.4 | 20.5 | 35.7 | 25.0 | **39.3** |
| 22 | 35.7 | 38.4 | 17.9 | 43.5 | **57.1** |
| 23 | 32.1 | **33.9** | 21.4 | 32.1 | 25.0 |
| 24 | 38.9 | 45.6 | 46.4 | 35.7 | **57.1** |
| 25 | 32.1 | 42.7 | 35.7 | 45.2 | **46.4** |
| 26 | 36.5 | 51.9 | 42.9 | 53.6 | **60.7** |
| 27 | 57.1 | 44.3 | **60.7** | 53.6 | 50.0 |
| 28 | 39.3 | 42.7 | 50.0 | **53.6** | 53.6 |
| 29 | 50.8 | 26.4 | 42.9 | 46.4 | **64.3** |
| 30 | 32.1 | 30.9 | 25.0 | 14.3 | **46.4** |
| 31 | 57.1 | 60.4 | 60.7 | 56.0 | **64.3** |
| 32 | **53.6** | 39.8 | 32.1 | 35.0 | 41.9 |
| 33 | 35.7 | 33.9 | 32.9 | 17.9 | **42.9** |
| 34 | 14.3 | 21.2 | **27.0** | 10.7 | 17.9 |
| 35 | 46.4 | 28.0 | 39.3 | 35.7 | **50.0** |
| 36 | 14.3 | 12.7 | 20.8 | 35.7 | **39.3** |
| 37 | 28.6 | 26.4 | 35.7 | 32.1 | **40.6** |
| 38 | 25.0 | **56.1** | 50.0 | 21.4 | 50.0 |
| 39 | 25.0 | 19.2 | 28.6 | 35.7 | **42.9** |
| 40 | 39.3 | 32.0 | 42.9 | 32.1 | **54.9** |
| 41 | 10.7 | **38.4** | 28.6 | 14.3 | 24.5 |
| 42 | 46.4 | 41.3 | 18.2 | 26.0 | **53.6** |
| 43 | 32.1 | 16.0 | 28.6 | 35.7 | **64.3** |
| 44 | 31.4 | 33.2 | 20.2 | 35.7 | **50.0** |
| 45 | 25.0 | **35.5** | 17.9 | 21.4 | 21.4 |
| 46 | 28.6 | 33.9 | 25.0 | **35.7** | 32.1 |
| 47 | 25.0 | 38.2 | 34.8 | 36.4 | **50.0** |
| 48 | 42.9 | 11.7 | 39.3 | 28.6 | **57.1** |
| 49 | 25.0 | 24.3 | 28.6 | 35.7 | **42.9** |
| 50 | 32.1 | 42.7 | **46.4** | 24.4 | 32.1 |
| 51 | 17.9 | **47.3** | 32.1 | 21.4 | 33.5 |
| 52 | 21.4 | **44.3** | 21.4 | 39.3 | 44.3 |
| 53 | 17.9 | 11.7 | 28.6 | 22.7 | **35.7** |
| 54 | 32.1 | 39.7 | 42.9 | 35.7 | **45.2** |
| 55 | 53.6 | 34.7 | **57.1** | 39.3 | 46.4 |
| 56 | 42.9 | 33.9 | 32.1 | 41.3 | **46.4** |
| 57 | 35.7 | 35.5 | 35.7 | 37.4 | **50.0** |
| 58 | 42.9 | 30.9 | 45.4 | 32.1 | **50.0** |
| 59 | 35.7 | **47.2** | 28.6 | 39.3 | 35.7 |
| 60 | **39.3** | 28.0 | **39.3** | 25.0 | 35.7 |
| 61 | **46.4** | 39.7 | 39.3 | 44.2 | 39.3 |
| 62 | 21.4 | 25.5 | **32.1** | 25.0 | 27.1 |
| 63 | 32.1 | 39.8 | 35.1 | **46.4** | 39.3 |
| 64 | 53.6 | 59.0 | 46.4 | 42.9 | **60.7** |
| 65 | 28.6 | 42.7 | 32.1 | 32.1 | **47.1** |
| 66 | 28.6 | **53.1** | 14.3 | 28.6 | 32.1 |
| 67 | 35.7 | 31.0 | 35.7 | 35.7 | **57.1** |
| 68 | 21.4 | 24.8 | 25.0 | 10.7 | **42.9** |
| 69 | 42.1 | 20.5 | 32.1 | 31.0 | **57.1** |
| 70 | 35.4 | **50.1** | 39.3 | 42.9 | 46.4 |
| 71 | 35.7 | 33.9 | 28.6 | 28.6 | **50.0** |
| 72 | **36.0** | 35.5 | 32.1 | 35.7 | 25.0 |
| 73 | 25.0 | 21.2 | **28.6** | **28.6** | 25.0 |
| 74 | 25.0 | 25.1 | 21.4 | **39.3** | 17.9 |
| 75 | 21.4 | **50.1** | 17.9 | 50.0 | 32.1 |
| 76 | 34.9 | **50.2** | 35.1 | 11.1 | 50.0 |
| 77 | 49.0 | 30.9 | 46.4 | **53.6** | 50.0 |
| 78 | 39.3 | 33.9 | 26.5 | 25.0 | **46.4** |
| 79 | **53.6** | 28.5 | 25.0 | 32.1 | 39.3 |
| Avg | 35.3 | 35.5 | 34.5 | 34.3 | **45.7** |
| Std | 11.1 | 11.8 | 10.4 | 10.5 | 11.8 |
| $t$(78) | 8.22 | 6.24 | 8.40 | 8.30 | - |
| Bonferroni corrected $p$ | <0.001 | <0.001 | <0.001 | <0.001 | - |

*We reported t-values and Bonferroni corrected p-values in the paired t-tests between the performance of CLISA and other models.*

TABLE S5
THE CLASSIFICATION ACCURACIES FOR EACH SUBJECT IN THE SEED DATASET

| Subject No. | Cross-subject emotion recognition | | | | | Generalizability test | | | | |
|---|---|---|---|---|---|---|---|---|---|---|
| | DE+MLP | SA | Corr-CA | Seq-CLR | CLISA | DE+MLP | SA | Corr-CA | SeqCLR | CLISA |
| 1 | **80.3** | 75.5 | 59.8 | 74.0 | 75.7 | **71.8** | 57.4 | 63.6 | 36.7 | 66.6 |
| 2 | 80.4 | 77.2 | 69.2 | 73.7 | **82.2** | 68.4 | 75.2 | 55.4 | **76.3** | 62.3 |
| 3 | 88.2 | **92.6** | 76.4 | 77.9 | 87.6 | 85.4 | 83.7 | 92.8 | 86.7 | **94.4** |
| 4 | 79.3 | 70.3 | 71.7 | 69.7 | **85.8** | 69.6 | 61.8 | 66.8 | 56.3 | 64.6 |
| 5 | 80.6 | 75.7 | 71.3 | **87.3** | 86.0 | 80.0 | 83.7 | 71.9 | **90.0** | 82.0 |
| 6 | 75.9 | 83.3 | 79.7 | 80.1 | **86.6** | **83.4** | 78.8 | 63.1 | 80.9 | 67.6 |
| 7 | 69.7 | 77.3 | 76.0 | 66.6 | **88.6** | 68.6 | 88.4 | 77.8 | 62.2 | **92.3** |
| 8 | 85.4 | 81.0 | 81.2 | 85.2 | **93.2** | 85.4 | 79.6 | **89.6** | 82.8 | 88.5 |
| 9 | 75.5 | 65.4 | 69.6 | 79.0 | **90.8** | 57.0 | **67.1** | 62.4 | 48.5 | 65.5 |
| 10 | 78.6 | 77.0 | 67.9 | **84.2** | 83.4 | 64.8 | 79.0 | **81.9** | 78.1 | 74.0 |
| 11 | **95.8** | 85.1 | 94.5 | 90.2 | 91.7 | 90.1 | 72.2 | 58.0 | 73.2 | **94.3** |
| 12 | 59.4 | **72.6** | 57.7 | 55.3 | 69.8 | 54.1 | **63.7** | 47.2 | 44.7 | 60.9 |
| 13 | 71.9 | 76.5 | 59.6 | 78.0 | **91.7** | 54.9 | **64.7** | 59.3 | 60.6 | 62.0 |
| 14 | 87.5 | 83.4 | 86.8 | 85.3 | **90.3** | 87.8 | 81.2 | 88.8 | **97.8** | 94.1 |
| 15 | 90.5 | 76.9 | 84.4 | 89.3 | **93.0** | **94.3** | 72.4 | 84.7 | 91.8 | 91.7 |
| Avg | 79.9 | 78.0 | 73.7 | 78.4 | **86.4** | 74.4 | 73.9 | 70.9 | 71.1 | **77.4** |
| Std | 8.7 | 6.3 | 10.2 | 9.2 | 6.4 | 12.8 | 8.9 | 13.6 | 18.0 | 13.4 |
| $t$(14) | 3.37 | 4.10 | 6.32 | 4.63 | - | 1.28 | 1.30 | 2.50 | 1.71 | - |
| Bonferroni corrected $p$ | 0.046 | 0.011 | <0.001 | 0.004 | - | 1.000 (0.221) [a] | 1.000 (0.214) | 0.260 | 1.000 (0.109) | - |

*We reported t-values and Bonferroni corrected p-values in the paired t-tests between the performance of CLISA and other models.*
*[a] The p-values in parentheses were uncorrected.*